\documentclass[acmsmall,screen,noacm=true]{acmart}
\setcopyright{none}
\renewcommand\footnotetextcopyrightpermission[1]{}
\makeatletter
\def\@copyrightspace{\relax}
\makeatother

\setcopyright{none}
\fancypagestyle{plain}{
  \fancyhf{} 
}

\author[Violet Szabó]{Violet Szabó} \affiliation{
  \institution{TU Delft}
  \country{Netherlands}
}
\email{V.Szabo-2@student.tudelft.nl}

\author[Dominik Winterer]{Dominik Winterer} \affiliation{
  \institution{ETH Zurich}
  \department{Department of Computer Science}
  \country{Switzerland}
}
\email{dominik.winterer@inf.ethz.ch}

\author{Zhendong Su}
\affiliation{
  \institution{ETH Zurich}
  \department{Department of Computer Science}
  \country{Switzerland}
}
\email{zhendong.su@inf.ethz.ch}

\newcommand{\algorithmOne}{SampleProgramInterval}
\newcommand{\algorithmTwo}{EstimateIndex}
\newcommand{\cCQ}{48.11}
\newcommand{\cCQPrecise}{48.110}
\newcommand{\erlangCQ}{6.51}
\newcommand{\erlangCQPrecise}{6.511}
\newcommand{\csharpCQ}{1.69}
\newcommand{\csharpCQPrecise}{1.691}
\newcommand{\cppCQ}{0.60}
\newcommand{\cppCQPrecise}{0.598}
\newcommand{\kotlinCQ}{0.31}
\newcommand{\kotlinCQPrecise}{0.308}
\newcommand{\javaCQ}{0.27}
\newcommand{\javaCQPrecise}{0.265}
\newcommand{\haskellCQ}{0.13}
\newcommand{\haskellCQPrecise}{0.128}

\newcommand{\fortranCQPrecise}{0.033}

\newcommand{\cobolCQPrecise}{0.032}
\newcommand{\goCQ}{0.03}
\newcommand{\goCQPrecise}{0.030}
\newcommand{\swiftCQ}{0.02}
\newcommand{\swiftCQPrecise}{0.018}
\newcommand{\rustCQ}{0.00}
\newcommand{\rustCQPrecise}{0.0004}

\newcommand{\sampleTarget}{100,000}
\newcommand{\sampleTargetJavaKotlin}{10,000}

\newcommand{\ie}{\hbox{\emph{i.e.}}\xspace}
\newcommand{\eg}{\hbox{\emph{e.g.}}\xspace}
\newcommand{\etc}{\hbox{\emph{etc.}}\xspace}

\newcommand{\wrt}{\hbox{\emph{w.r.t.}}\xspace}
      
\newcommand{\pl}{\mathit{PL}}
\newcommand{\cqSymbol}[1]{\mathsf{\bf CQ}(\text{\emph{#1}})}
\newcommand{\maxbytes}{256}
\newcommand{\expReptitions}{three}
\newcommand{\TransformationScriptLines}{1,000}
\newcommand{\cfggrammar}{\ensuremath{G}}
\newcommand{\toolname}{\textsf{cq-test}}

\newcommand{\nonterminals}{\ensuremath{N}}
\newcommand{\terminals}{\ensuremath{\Sigma}}
\newcommand{\productions}{\ensuremath{P}}
\newcommand{\startSymbol}{\ensuremath{S}}

\newcommand{\rtggrammar}{\ensuremath{G'}}
\newcommand{\rankedAlphabet}{\ensuremath{\Sigma'}}

\newcommand{\sampleSet}{\ensuremath{S}}
\newcommand{\plEnumeration}{\ensuremath{f_{\pl}}}
\newcommand{\sampleSize}{\ensuremath{n}}
\newcommand{\sampleMinLength}{\ensuremath{a}}
\newcommand{\sampleMaxLength}{\ensuremath{b}}
\newcommand{\featMinIndex}{\ensuremath{i_a}}
\newcommand{\featMaxIndex}{\ensuremath{i_b}}
\newcommand{\featMinIndexAppr}{\ensuremath{i_a'}}
\newcommand{\featMaxIndexAppr}{\ensuremath{i_b'}}
\newcommand{\indexSet}{\ensuremath{I}}
\newcommand{\oversamplingFactor}{\ensuremath{\alpha}}
\newcommand{\boundGrowFactor}{\ensuremath{\beta}}

\newcommand{\maxbytesC}{1000}
\newcommand{\windowSizeSymbol}{\ensuremath{\varepsilon}}
\newcommand{\windowRadius}{\ensuremath{5}}

\newcommand{\maxCnt}{step\_increase\_threshold}
\newcommand{\step}{step}
\newcommand{\cnt}{steps\_taken}
\newcommand{\length}[1]{size(#1)}
\newcommand{\best}{best}
\newcommand{\curr}{curr}
\newcommand{\start}{start}
\newcommand{\indexEnd}{end}
\newcommand{\maxStep}{max\_tries}
\newcommand{\maxByLength}[2]{max\_by\_size(#1, #2)}

\usepackage{amsfonts}
\usepackage{amsmath}
\usepackage{mathtools}
\usepackage{array}
\usepackage{graphicx}
\usepackage{DotArrow}
\usepackage{balance}
\usepackage{microtype}
\usepackage{xcolor}
\usepackage{wrapfig}
\usepackage{fancyvrb}
\usepackage{adjustbox}

\usepackage[frozencache]{minted}
\setminted{breaklines, fontsize=\footnotesize, breaksymbol={}}
\AtBeginEnvironment{minted}{%
  }

\newcolumntype{R}[2]{%
    >{\adjustbox{angle=#1,lap=\width-(#2)}\bgroup}%
    l%
    <{\egroup}%
}

\colorlet{shadecolor}{gray!10}

\usepackage{xspace}
\usepackage{algorithm}
\usepackage[noend]{algpseudocode}
\usepackage{setspace}

\usepackage{xpatch}
\makeatletter
\xpatchcmd{\algorithmic}{\itemsep\z@}{\itemsep=1.1pt}{}{}
\makeatother

\usepackage{pifont}
\usepackage{DotArrow}
\usepackage{subcaption}
\usepackage[tableposition=top]{caption}
\usepackage{tikz}
\usetikzlibrary{external}
\usetikzlibrary{arrows,automata}
\usetikzlibrary{shapes.misc, positioning}
\usetikzlibrary{patterns}
\usepackage{pgfplots}
\usetikzlibrary{matrix}
\pgfplotsset{compat=newest}
\usepgfplotslibrary{groupplots}

\usepackage{enumitem}
\setlist[description]{font=\bfseries}
\usepackage{colortbl}
\usepackage{tabularx}
\usepackage{graphbox}
\usepackage{longfbox}
\usepackage{booktabs}
\usepackage{listings}
\newcommand{\lstbg}[3][0pt]{{\fboxsep#1\colorbox{#2}{\strut #3}}}
\lstdefinelanguage{diff}{
  basicstyle=\ttfamily\small,
  morecomment=[f][\lstbg{gray!30}]-,
  morecomment=[f][\lstbg{gray!15}]+,
  morecomment=[f][\textit]{@@},
}
\fboxset{padding={1.0ex,1.0ex, 1.0ex, 1.0ex}, margin={1em,0.0em,0em}}
\newtheorem{innercustomgeneric}{\customgenericname}
\providecommand{\customgenericname}{}
\newcommand{\newcustomtheorem}[2]{%
  \newenvironment{#1}[1]
  {%
   \renewcommand\customgenericname{#2}%
   \renewcommand\theinnercustomgeneric{##1}%
   \innercustomgeneric
  }
  {\endinnercustomgeneric}
}
\newcustomtheorem{customtheorem}{Proposition}
\newcustomtheorem{customlemma}{Lemma}
\newtheorem{definition}{Definition}

\newcommand{\grammarStats}{
\footnotesize
\begin{tabular}{lcrrr}
\toprule
\textbf{Grammar} &  \textbf{\# nont.} & \textbf{\# ter.} & \textbf{\# prod.} \\
C & 83 & 118 & 163 \\
C++ & 188 & 141 & 305 \\
Java & 123 & 122 & 213 \\
C\# & 207 & 162 & 322 \\
Fortran & 335 & 177 & 377 \\
Go & 103 & 70 & 115 \\
Swift & 307 & 170 & 320 \\
Rust & 194 & 108 & 407 \\
Kotlin & 150 & 142 & 163 \\
COBOL & 590 & 558 & 280 \\
Haskell & 223 & 132 & 248 \\
Erlang & 122 & 8 & 129 \\
\bottomrule
\end{tabular}
}

\begin{document}
\title[CQ: A Metric for Compilation Hardness of Compiled PLs]
{Compilation Quotient (CQ): A Metric for the Compilation Hardness of Programming Languages}
\begin{abstract}
Today's programmers can choose from an exceptional range of programming languages,  
each with its own traits, purpose, and complexity. A key aspect of a language's 
complexity is how hard it is to compile programs in the language. While most 
programmers have an intuition about compilation hardness for different 
programming languages, no metric exists to quantify it. We introduce the compilation 
quotient (CQ), a metric to quantify the compilation hardness of compiled programming 
languages. The key idea is to measure the compilation success rates of programs 
sampled from context-free grammars. To this end, we fairly sample over 12 million programs 
in total. CQ ranges between 0 and 100, where 
0 indicates that no programs compile, and 100 means that all programs compile. 
Our findings on 12 popular compiled programming languages show high variation 
in CQ.  C has a CQ of \cCQ{}, C++ has \cppCQ{}, Java has \javaCQ{} and 
Haskell has \haskellCQ{}. Strikingly, Rust's CQ is nearly 0, and for C, even a 
large fraction of very sizable programs compile. We believe CQ can help understand 
the differences of compiled programming languages better and help language designers.   
\end{abstract}

\maketitle    
\section{Introduction}
Programming languages profoundly influence how we build software. Each programming 
language has its own characteristic traits, purposes, infrastructure, 
and complexity. One key aspect of a language's complexity is how hard 
it is to produce \emph{valid} programs $P$ in the language. If the language is compiled, 
an ahead-of-time compiler decides whether $P$ is valid, produces an executable in case of success, 
and rejects the program with an error otherwise. Rejected programs frustrate 
programmers, leading to reduced productivity, failed or abandoned projects, \etc 
While most programmers have an intuition of the hardness of writing compiling programs 
in different programming languages, no metric exists to quantify it. Without a 
metric, language designers rely solely on user experiences, which can be valuable          
but are also potentially biased and contradictory. Moreover, without a metric, learning from 
historical languages' benefits and flaws is difficult. However, realizing a metric 
to quantify the compilation hardness of different languages is challenging since it 
requires inferring compilation hardness on the (infinite) space of all programs.  

{\parindent0pt 
\paragraph{\textbf{Compilation Quotient}}
We propose the \emph{compilation quotient (CQ)}, a metric for the compilation hardness 
of programming languages. The key idea is to measure CQ for programming 
language $\pl$ by sampling programs from context-free grammars for $\pl$. Concretely, 
the CQ of $\pl$ is the percentage of accepted programs over the total number of sampled 
programs for $\pl$. CQ ranges between 0 and 100, where 0 indicates that no 
programs compile, and 100 means that all programs compile. We compute the 
CQ for a given language $\pl$ by the following steps: (1) translate the context-free 
grammar into algebraic datatypes, (2) sample a large number of programs using 
off-the-shelf 
property-based tester FEAT, (3) forward each of these programs to a $PL$ compiler 
and track the results, and (4) compute the CQ by dividing the number of 
compiling samples by the number of total samples.  We obtain the context-free 
grammars for the languages from the repository of ANTLR, the 
parser generator~\cite{parr-etal-oopsla2014}. To sample programs of diverse sizes, 
we use a bucket-based algorithm with lower and upper bounds on program size 
(see Section~\ref{subsec:computing}). We use the most popular 
compiler for each language, e.g., \texttt{\small gcc} for C, \texttt{\small g++} 
for C++, \texttt{\small rustc} for Rust, and 
\texttt{\small ghc} for Haskell. 
}

\begin{figure}
    \centering
    \begin{subfigure}[b]{0.38\textwidth}
        \centering
        \footnotesize
        \begin{tabular}{p{0.6cm}p{1.8cm}r}
            \toprule
            \textbf{\#} &  \textbf{Language} &  \textbf{CQ} \\
            1 & C & \cCQPrecise\ \\
            2 & Erlang & \erlangCQPrecise\ \\
            3 & C\# & \csharpCQPrecise\ \\
            4 & C++ & \cppCQPrecise\ \\
            5 & Kotlin & \kotlinCQPrecise\ \\
            6 & Java & \javaCQPrecise\ \\
            7 & Haskell & \haskellCQPrecise\ \\
            8 & Fortran & \fortranCQPrecise\ \\
            9 & COBOL & \cobolCQPrecise\ \\
            10 & Go & \goCQPrecise\ \\
            11 & Swift & \swiftCQPrecise\ \\
            12 & Rust & \rustCQPrecise\ \\
            \bottomrule
            \end{tabular}
        \caption{\label{a}}
    \end{subfigure}
    \hspace{0.3cm}
    \begin{subfigure}[b]{0.55\textwidth}
        \includegraphics[width=1.0\textwidth]{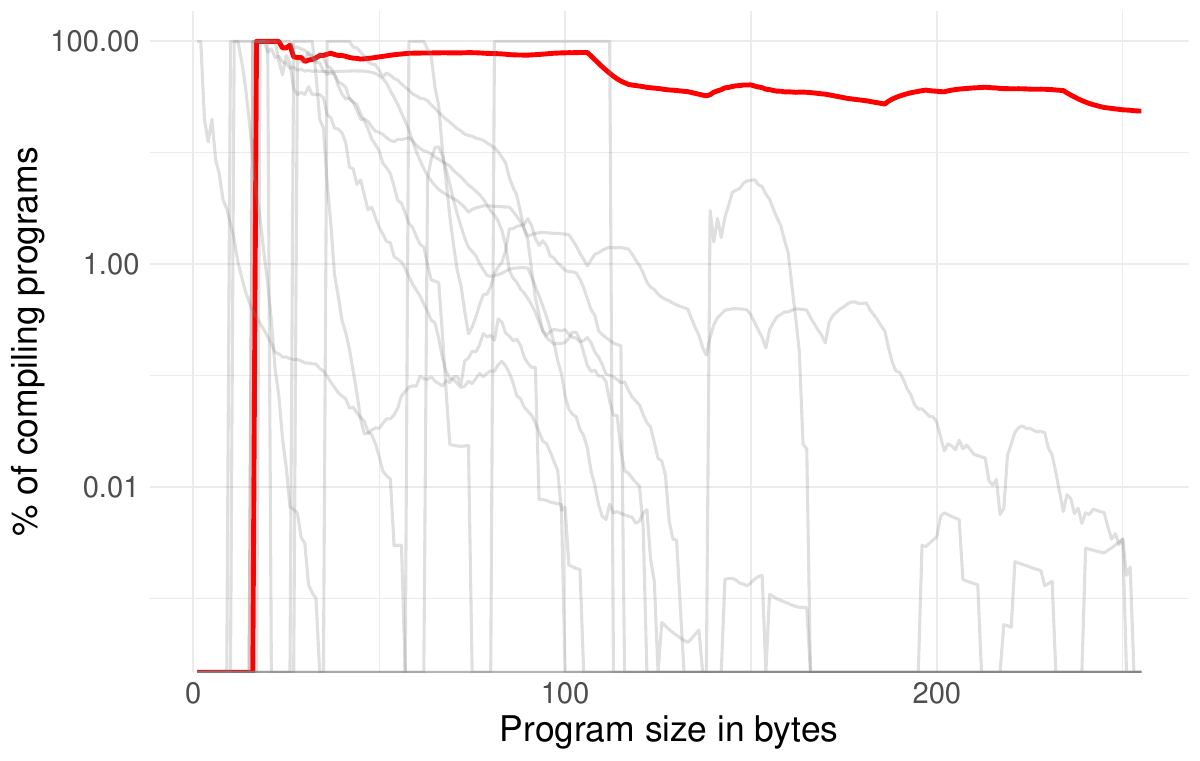}
        \caption{\label{b}}
    \end{subfigure}
    \caption{(a) CQ-ranking of twelve popular compiled languages, 
    and (b) C's divergence.}
\end{figure}

{\parindent0pt 
\paragraph{\textbf{Results \& Insights}}
We compute the CQs of twelve popular compiled programming languages. 
Our findings show high variation in CQ (see Fig.~\ref{a}). The language with the 
highest CQ is C. It has a CQ of \cCQ{} leading by a large margin over all other languages. 
C++ has a CQ of \cppCQ{}, Java has a CQ of \javaCQ{} and Haskell has a CQ of \haskellCQ{}. 
Other interesting observations include that the most popular object-oriented programming 
languages all rank high in the CQ ranking (\ie, C\#, Java, and C++). Besides, we also investigate 
the percentage of valid programs as program size increases (see Fig.~\ref{b}). Strikingly, 
for C, the percentage of valid C programs converges to a strictly 
positive value as program size increases, unlike the distributions of 
all other languages which drop to zero instead. Complementing quantitative 
results, we analyze sample programs from each language, identify notable features, 
and discuss their effect on CQ, presenting patterns between different languages.
}

{\parindent0pt 
\paragraph{\textbf{Main Contributions}}
We make the following main contributions: 
\begin{itemize}
\setlength\itemsep{0.25em}
\item We introduce the \emph{compilation quotient (CQ)}, a metric for the compilation 
hardness of programming languages;  
\item We propose a framework to compute CQs, including a sampling algorithm 
and a practical tool \toolname{} for sampling more than 12 million programs;     
\item We calculate CQ for twelve popular compiled programming languages including 
C, C++, Java, C\#, Kotlin, Haskell, and Rust;  
\item We conduct in-depth analyses on the distribution of valid programs \wrt program size, 
     how various language features affect CQ, discuss the meaning of the metric, \etc 
\end{itemize}

We will open-source source \toolname{} and plan to submit it to artifact evaluation.  
Moreover, \toolname{} and the experimental data is attached to this submission.
}

{\parindent0pt 
\paragraph{\textbf{Structure of the paper}}
The paper realizes an empirical study. First, we describe its setup 
(Section~\ref{sec:setup}) including programming languages, sampling \etc, then 
we describe the results (see Section~\ref{sec:result}) and interpret them  
(Section~\ref{sec:posthoc}). We discuss the implications of CQ 
(Section~\ref{sec:discussion}), survey 
related work (Section~\ref{sec:related-work}), and close with a conclusion 
(Section~\ref{sec:conclusion}). 
}

\section{Empirical Setup}
\label{sec:setup}
We first describe our selection of compiled programming languages 
(Section~\ref{subsec:chosenlangs}) including their context-free grammars 
(Section~\ref{subsec:grammars}). Second, we introduce the compilation 
quotient (CQ), a metric for the compilation hardness of programming languages 
(Section~\ref{subsec:cq}). Finally, we describe algorithms for computing   
CQ (Section~\ref{subsec:computing}) and describe additional setup details (Section~\ref{subsec:further-setup}). 
\subsection{Compiled Programming Languages}
\label{subsec:chosenlangs}
We consider \emph{compiled programming languages} satisfying the following 
three criteria: (1) the language's source code is written in a human-readable, 
high-level language,(2) the language's source code is transformed by an ahead-of-time compiler to realize 
an executable\footnote{Either as machine code or an intermediate representation. } and (3) the executable can be executed directly on a computer or indirectly by another runtime system. For this work, we restrict ourselves to compiled languages 
as they have a clear distinction between semantic errors and runtime errors, 
unlike interpreted languages or hybrids. We make a selection of compiled programming 
languages including those in the top 20 of the TIOBE index (see Fig.~\ref{tiobe}), 
the leading language popularity metric. For a programming language $\pl$, the index can be 
thought of as the percentage of $\pl$'s weighted hits \wrt the total weighted search hits 
of all programming languages in major  web search engines including Google, Bing, 
Baidu, Wikipedia, and YouTube. C, for example, has the highest TIOBE index (11.27\%) which 
puts it at the top of the ranking. Besides the compiled languages in the top 20 of 
TIOBE, we consider Haskell because of its importance to the PL community, and COBOL and Erlang for historical importance. We exclude non-textual languages.  
\begin{figure*}
\begin{subfigure}[b]{0.45\textwidth}
\centering
\footnotesize
\begin{tabular}{llr}                                                          
\toprule
\textbf{\#} &  \textbf{Language} &  \textbf{Score} \\
2 & C & 11.27\% \\
3 & C++ & 10.65\% \\
4 & Java & 9.49\% \\
5 & C\# & 7.31\% \\
\textcolor{gray}{7} & \textcolor{gray}{Visual Basic} & \textcolor{gray}{2.22\%} \\
11 & Fortran & 1.28\% \\
12 & Go & 1.19\% \\
\textcolor{gray}{15} & \textcolor{gray}{Delphi/Object Pascal} & \textcolor{gray}{1.02\%} \\
16 & Swift & 1.00\% \\
17 & Rust & 0.97\% \\
20 & Kotlin & 0.90\% \\
21 & COBOL & 0.88\% \\
28 & Haskell & 0.65\% \\
51 - 100 & Erlang & < 0.24\% \\
\bottomrule
\end{tabular}
\caption{\label{tiobe}}
\end{subfigure}
%
%
\begin{subfigure}[b]{0.45\textwidth}
\centering
\grammarStats
\caption{\label{grammar-stats}}
\end{subfigure}
\caption{(a) Selected TIOBE score of programming languages with rank (\#) as of September 2023. 
(b) ANTLR grammar statistics: nonterminals (\# nont.), terminals (\# ter.) and production 
counts (\# prod.). Grayed: For Visual Basic and Delphi/Object Pascal there were 
no suitable ANTLR grammars.       
\label{language-table}}
\end{figure*}
%


\subsection{Grammars for Programming Languages}
\label{subsec:grammars}
For each chosen programming language $\pl$, we use a context-free grammar from the ANTLR grammar 
repository.~\footnote{\url{https://github.com/antlr/grammars-v4}} If there are 
multiple grammars for a programming language $\pl$, we choose the grammar with the newest 
version of $\pl$. We aim at following $\pl$'s official language specification as closely      
as possible, without sampling too many equivalent programs.
To that end, we modify each grammar as follows: (1) we bound the number  
of identifiers (\ie variables) to two, (2) we restrict the number of literals 
per datatype to one, \eg, "string" for strings, "$42$" for integers, "$123.4$" 
for doubles \etc, except for booleans for which we allow both "true" and "false", 
(3) we create a single entry point, usually a main function, with all statements, and (4) we
fix orders of modifiers to avoid combinatorial explosions. Besides these changes, we describe the 
following further changes:   

\begin{description}
\item[C.] The grammar for C realizes the C11 standard~\cite{c11-standard}. The 
grammar includes non-standard extensions, and datatypes for processors supporting
Single Instruction Multiple Data (SIMD).  We remove all these extensions and 
additionally remove inline assembly.\\ 
\item[C++.] The grammar for C++ realizes the C++14 standard~\cite{cpp14-standard}.
We remove attributes, user-defined literals, and alternative operator representations.
Similar to C, we remove inline assembly. \\ 

\item[C\#.] The grammar for C\# realizes version 6 of the language~\cite{csharp6-docs}. 
We remove string interpolation.  Furthermore, we remove unsafe modifiers as these never 
compile without special compiler flags. Likewise, we remove pointer operations since 
these are only valid in an unsafe context.\\  

\item[Fortran.] The Fortran grammar realizes the 1990 standard of the language~\cite{fortran90}.  
We reduce duplicate programs by forcing uppercase keywords, separable
keywords to be written together (\eg \texttt{\small GOTO} instead of \texttt{\small GO TO}),
and operators to be in their text representation (\eg \texttt{\small .NE.} instead of \texttt{\small /=}).
We restrict labels to two values ($100$ and $200$) and restrict literals to those supported by
GFortran. We remove \texttt{\small IMPLICIT} statements except for \texttt{\small IMPLICIT NONE}.~\footnote{\texttt{\small IMPLICIT}
statements dictate the type of variables with names starting with a given letter but no explicit type. However
all our generated variables start with the letter 'v'. \texttt{\small IMPLICIT\ NONE} disables implicit typing.}
We further remove import statements for external files. Finally,
we remove format statements.
\\ 

\item[Java, Go.] For the Java and Go grammars, no language-specific modifications were needed.
The Java grammar realizes the Java 17 ~\cite{java-spec-17} and the Go grammar 
realizes version 1.20~\cite{golang120}. \\ 

\item[Swift.] The Swift grammar realizes version 5.4 \cite{swift54}. We remove 
extended string literals, imports, availability conditions, conditional compilation statements, and attributes.
We restrict the language to the plus and minus operators to avoid combinatorial explosion. 
We restrict implicit parameter names to \$0 and \$1. As the language has many modifiers,
only valid for a specific declaration type, we tie modifiers to the correct
construct.\\

\item[Rust.]  The Rust grammar realizes version v1.60.0~\cite{rust-160}. We 
remove macros and attributes from the grammar. We also remove the \texttt{\small extern} modifier for functions. 
\\

\item[Kotlin.] The grammar for Kotlin realizes version 1.4~\cite{kotlin-14}.
We remove annotations, fix the order of modifiers, and tie modifiers to the correct
declaration type. \\

\item[COBOL.] The grammar realizes COBOL 85 \cite{cobol85standard}. We removed 
null-terminated string literals, as well as CICS and SQL extensions.
We define the entry point as a procedure division, optionally proceeded by a data division
that may contain only a working-storage section. Only level numbers $01$, $02$, $66$, $77$ and $88$
are allowed in the working-storage section. To avoid duplicate programs, we force 
operators to use their character form instead of the more verbose syntax (\eg $x > y$ 
instead of $x \texttt{\small GREATER THAN  } y$).
We force the usage of \texttt{\small PICTURE} instead of \texttt{\small PIC}.
\\

\item[Haskell.] The grammar for Haskell realizes the 2010 version of the language~\cite{haskell-2009}. 
We remove pragmas and qualified variable symbols. We remove constructor symbols as they can never be defined
in programs generated by our setup. We restrict operator symbols to the plus and minus symbols, as otherwise,
the number of permitted identifiers would be very large.
As the entry point, we use: 
\begin{lstlisting}[language=Haskell,numbers=none]
main = return ()
func = ...  where { varX = ...,... }
\end{lstlisting}
We permit \texttt{\small func} to be used as an identifier to enable generation of recursive 
programs. 
\\

\item[Erlang.] The grammar realizes Erlang 23.3~\cite{erlang}.
The only language-specific modification made was to reduce the number of atom literals to one,
\ie, \texttt{\small an\_atom}.
\end{description}
For Visual Basic (VB), there is only an ANTLR grammar for an outdated version 6.0 
released in 1998. Compilers for this version were only available for Windows with an 
enterprise subscription. Moreover, for Delphi/Object Pascal there were no 
ANTLR grammar is available at all.  We hence do not consider these two languages for our   
study.
%
\subsection{Compilation Quotient}
\label{subsec:cq}
With the language grammars specified, we move on to the compilation quotient. 
First, we need a few basic, formal definitions, then present CQ and LCQ. 

{\parindent0pt 
\paragraph{\textbf{Definitions}} We assume familiarity with context-free grammars (CFG) including 
derivations, terminals, nonterminals \etc Let $\pl$ be a programming language, 
then we describe with $G_\pl$ the associated context-free grammar of $\pl$. 
We denote the filesize of program $P$ (in bytes) by $\mathit{size}(P)$. 
The induced language of a grammar $L(G_\pl)$ is the set of all words, \ie, programs, 
derivable from $G_\pl$. For the \emph{semantically correct} programs $P$ in $L(G_\pl)$,    
we write $\vdash P$. We further define $L_S(G_\pl) = \{ P \in L(G_\pl) \leq S \}$    
as the bounded language of $G_\pl$ for a fixed size bound $S$.}

\noindent\begin{definition}[Compilation Quotient]
The compilation quotient (CQ) is defined as follows: 
\[
\mathsf{\bf CQ}(\pl) = \frac{|\{ P \in L_S(G_\pl) \mid \ \vdash P \}|}{|\{ P \in L_S(G_\pl) \}| } \times 100
\]

where $\pl$ is a programming language, $G_\pl$ is a CFG for $\pl$, and $S$ is a fixed size bound. 
\end{definition}

The intuition behind CQ is measuring the semantic constraints that a given         
PL imposes on the programmer. CQ ranges between $0$ and $100$. A CQ of $0$          
for a language $\pl$ indicates that all programs in $L(G_\pl)$ are semantically invalid. A CQ of  
$100$, on the other hand, means that all programs are semantically valid. In practice, 
we use the feedback of a compiler of $\pl$ to determine the semantic validity 
of $\pl$. For the computation of CQ, we will further assume programs of  
$\mathit{size}(P)$ less than \maxbytes{} bytes. The CQ is a single number 
describing a given language as a whole. To analyze the distribution of programs, 
we introduce the local compilation quotient:
\noindent\begin{definition}[Local Compilation Quotient]
  The local compilation quotient (LCQ) is defined as follows:
  \[
      \mathsf{\bf LCQ}(\pl, x, \windowSizeSymbol) = \frac{|\{ P \in L(G_\pl) \wedge x - \windowSizeSymbol \leq size(P) \leq x + \windowSizeSymbol \mid \ \vdash P \}|}{|\{ P \in L(G_\pl) \wedge x - \windowSizeSymbol \leq size(P) \leq x + \windowSizeSymbol \}| } \times 100
  \]
where $\pl$ is a programming language, $G_\pl$ is a context-free grammar for 
$\pl$, $x$ is an integer indicating program size, and $\windowSizeSymbol$ is an integer indicating 
a window radius.
\end{definition}

The LCQ is intuitively a decomposition of the CQ for the given language, describing
the proportion of accepted programs at size $x$ and smoothing factor $\windowSizeSymbol$.  
For this paper, we will fix $\windowSizeSymbol = \windowRadius$.

\subsection{Computing Compilation Quotients} 
\label{subsec:computing}
We now describe how to compute CQ.
We first compile the context-free grammars to regular tree grammars,       
then describe the sampling algorithm along with an example.

{\parindent0pt 
\paragraph{\textbf{Compiling context-free grammars to regular-tree grammars.}} 
To sample programs from a context-free grammar, we compile the context-free grammar 
to a regular tree grammar, \ie, algebraic datatypes in a functional programming language such as 
Haskell.  
A \emph{context-free grammar}  $\cfggrammar = \langle \nonterminals, \terminals, \productions, \startSymbol \rangle$
consists of nonterminals $\nonterminals$, terminals $\terminals$, productions $\productions$, and    
a start $\startSymbol$ from $\nonterminals$. 
A \emph{regular tree grammar} 
$\rtggrammar = \langle \nonterminals', \rankedAlphabet, \productions', \startSymbol' \rangle$ 
consists of nonterminals $\nonterminals'$, a ranked alphabet $\rankedAlphabet$, productions 
$\productions'$ and a start symbol $\startSymbol'$. The elements in $\rankedAlphabet$ have  
constructors with arities where terminals are nullary. Intuitively, constructors of 
strictly positive arity represent tree patterns. Productions in $\productions'$        
have a single nonterminal on the left-hand side and symbols from $\nonterminals'$ and $\rankedAlphabet$ 
on their right-hand side. We compile a context-free grammar $\cfggrammar = \langle \nonterminals, 
\terminals, \productions, \startSymbol \rangle$  into a \emph{regular tree grammar} 
$\rtggrammar = \langle \nonterminals', \rankedAlphabet, \productions', \startSymbol' \rangle$ 
as follows: (1) we set $\nonterminals' = \nonterminals$ and $\startSymbol' = \startSymbol$, 
(2) we define $\rankedAlphabet$ as a superset of $\terminals$ with constructors 
$C^i_{\mathit{lhs}}(.)$ in $\rankedAlphabet$ for every production 
with left-hand side $\mathit{lhs}$, and (3) for $\productions'$, 
we append a constructor $C^i_{\mathit{lhs}}(.)$ to every production $p^i_{\mathit{lhs}}$ in $P$. 
The resulting regular-tree grammar is then converted into algebraic datatypes. 
We realize the compilation in about \TransformationScriptLines{} lines of Python code.
The script generates Haskell code for  exhaustive enumerator FEAT~\cite{duregard-etal-2012}. 
By this we obtain an enumeration of programs $\plEnumeration : \mathbb{N} \rightarrow L(G_\pl)$ 
as a mapping from natural numbers to programs of $\pl$.  
}

{\parindent0pt 
\paragraph{\textbf{Interval sampling of programs.}}
We sample programs from an interval bounded by a minimum $\sampleMinLength$ and maximum 
$\sampleMaxLength$ on filesize.  Given a sample size $\sampleSize$, we want to obtain samples from $S$.  
$$\sampleSet \subseteq \{P \in L(G_\pl)\ |\ \mathit{size}(P) \ \in [\sampleMinLength, \sampleMaxLength) \}
\hspace{0.4cm}
\indexSet =  \{i\ |\ \mathit{size}(\plEnumeration(i)) \in [\sampleMinLength, \sampleMaxLength) \}
$$
The set $\indexSet$ is an index set to generate elements in $S$.
Moreover, indices $\featMinIndex = \min \{ \indexSet \}$ and $\featMaxIndex = \max \{\indexSet\}$ 
locate the programs at the bounds. We want the indices of $\sampleSet$ to be evenly 
distributed within $\indexSet$, to accurately measure the distribution of programs 
within the interval. FEAT's  enumeration is monotonically increasing in the constructors. 
Montonicity is not guaranteed for filesize, \ie, for indices $i < j$  implies 
$\mathit{size}(\plEnumeration(i)) < \mathit{size}(\plEnumeration(j))$ does often 
hold but is not guaranteed. This implies that finding $\featMinIndex$ requires 
exhaustively checking every index up to $\featMinIndex$, which is infeasible. 
Hence we use the approximations $\featMinIndexAppr$ and $\featMaxIndexAppr$ 
for the boundaries of the interval. 
}
Our realization of the sampling algorithm is presented in Algorithm~\ref{algo:sample}.  
The algorithm takes $\sampleSize$, filesize bounds $a$ and $b$, and indices 
for $\featMinIndexAppr$ and $\featMaxIndexAppr$ as its input. We obtain  
$\featMinIndexAppr$ and $\featMaxIndexAppr$ from the procedure \textsc{\algorithmTwo} 
(see \ref{algo:bounds}). The procedure \textsc{\algorithmOne} returns the 
first index of a program of filesize $x$. We describe details on 
\textsc{EstimateIndex} in the next paragraph.
After initialization (Line \ref{SampleRange-Line1}), the algorithm enters 
its main loop in which it remains until either $\sampleSize$ samples were collected 
or it was unsucessful after $\mathit{max\_tries}$ attempts. For our 
experiments we set $\mathit{max\_tries}$ to 16. We next sample from the 
enumeration at evenly distributed indices between $\featMinIndexAppr$ and $\featMaxIndexAppr$.
We next remove programs that have an incorrect size (Line \ref{SampleRange-Line6}).
As this step reduces our candidate sample set, we oversample by a factor  
of $\oversamplingFactor$. We set $\oversamplingFactor$ to $8$ as our default 
configuration.  If there are exactly $\sampleSize$ samples left
after the filtering, the algorithm concludes. If there are too many samples, we randomly take
$\sampleSize$ without replacement (Line \ref{SampleRange-Line8}). If there are too few, we 
retry, adjusting $\featMinIndexAppr$ and $\featMaxIndexAppr$ using a factor 
$\boundGrowFactor$ (Line \ref{SampleRange-Line10}), and increasing our sampling
density to find more programs of the correct size. We set $\beta$ to be $2$.   
If \textsc{\algorithmOne} fails to find sufficient samples after $\mathit{max\_tries}$ 
attempts, it terminates returning the sample set. 

\begin{algorithm}[t]
  \begin{algorithmic}[1]
    \Function{\algorithmOne}{\sampleSize, \sampleMinLength, \sampleMaxLength, \featMinIndexAppr, \featMaxIndexAppr}
      \State $\best \gets \emptyset, \step \gets 0, \start \gets \featMinIndexAppr, \indexEnd \gets \featMaxIndexAppr$\label{SampleRange-Line1}
      \While{$|\best| < \sampleSize \land \step < \maxStep$}
        \State $\sampleSize ' \gets \oversamplingFactor \cdot \sampleSize \cdot (\step + 1) $
        \State $\curr \gets \{\plEnumeration(i)\: |\: \sampleSize ' \text{evenly spaced i's from [\start{}, \indexEnd{}) }\}$ \label{SampleRange-Line5}
        \State $\curr \gets \{w \in \curr\: |\: \sampleMinLength \leq \length{w} < \sampleMaxLength\}$ \label{SampleRange-Line6}
        \If{$|\curr| > \sampleSize$}
          \State $\curr \gets \text{take $\sampleSize$ elements from \curr{} without replacement}$ \label{SampleRange-Line8}
        \EndIf
        \State $\best \gets \maxByLength{\curr}{\best}$
        \State $\step \gets \step + 1, \start \gets \frac{\start}{\boundGrowFactor}, \indexEnd \gets \indexEnd * \boundGrowFactor$ \label{SampleRange-Line10}
      \EndWhile
      \State \Return $\best$
    \EndFunction
  \end{algorithmic}
  \caption{Sampling programs in a given size range}\label{algo:sample}
\end{algorithm}

\begin{algorithm}[t]
  \begin{algorithmic}[1]
  \Function{\algorithmTwo}{x}
  \State $i \gets 0, \cnt \gets 0, \step \gets 0$
    \While{$\length{\plEnumeration(i)} < x$} \Comment{Overestimate the index} \label{EstimateIndex-Line3}
      \If{$\cnt = \maxCnt$}
        \State $\step \gets \step + 1, \cnt \gets 0$ \label{EstimateIndex-Line5}
      \EndIf
      \State $i \gets i + 10^{\step}, \cnt \gets \cnt + 1$ \label{EstimateIndex-Line6}
    \EndWhile
    \If{$\length{\plEnumeration(i)} = x$}
      \State \Return $i$ \label{EstimateIndex-Line8}
    \EndIf
    \State $\cnt \gets 0, \step \gets 0$
    \While{$\length{\plEnumeration(i)} > x$} \Comment{Decrease the estimate} \label{EstimateIndex-Line10}
      \If{$\cnt = \maxCnt$}
        \State $\step \gets \step + 1, \cnt \gets 0$
      \EndIf
      \State $i \gets i - 10^{\step}, \cnt \gets \cnt + 1$ \label{EstimateIndex-Line13}
    \EndWhile
    \State \Return $i + 10^{\step}$ \label{EstimateIndex-Line14}
  \EndFunction
  \end{algorithmic}
  \caption{Returns an estimate on the first index of a program of filesize $x$}\label{algo:bounds}
\end{algorithm}

{\parindent0pt 
\paragraph{\textbf{Estimating index bounds.}}
\textsc{\algorithmOne{}} requires estimates for $\featMinIndexAppr$ and $\featMaxIndexAppr$. 
We use a exponential stepping algorithm in Algorithm \ref{algo:bounds} 
to determine $\featMinIndexAppr$. \textsc{\algorithmTwo{}} starts at zero and 
iterates through the natural numbers (Lines \ref{EstimateIndex-Line3}-\ref{EstimateIndex-Line6}), 
initially at a step size of one. After every $\mathit{\maxCnt}$ iterations 
$\mathit{step}$ is incremented by one. This multiples the gaps of consecutive     
accesses to \plEnumeration{} by ten. We do this because program size grows 
logarithmically with the index, thus a linear search would be inefficient.
Once an index $i'_1$ is reached such that
$size(\plEnumeration(i'_1)) \geq \sampleMinLength$, the step size is reset, and the
direction of search is reversed (Lines \ref{EstimateIndex-Line10}-\ref{EstimateIndex-Line13})
Once an index $i'_2$ is found such that
$size(\plEnumeration(i'_2)) < \sampleMinLength$, the algorithm returns the last 
index it encountered before $i'_2$ (Line \ref{EstimateIndex-Line14}). As a result, 
$\featMinIndexAppr$ will be a
slight overapproximation for $\featMinIndex$.
To determine $\featMaxIndexAppr$, we assume that $\featMaxIndex \approx
\min \{i \in \mathbb{N}\ |\ size(\plEnumeration(i)) = \sampleMaxLength\}$. This
lets us use the same algorithm to determine both $\featMinIndexAppr$ and
$\featMaxIndexAppr$.
}

{\parindent0pt 
\paragraph{\textbf{Avoiding bias towards longer programs.}}
\textsc{\algorithmTwo{}} is an effective approximation for sampling programs
sample within an interval $[\sampleMinLength,\sampleMaxLength)$. 
As the number of programs grows exponentially with size, however, \textsc{\algorithmTwo{}}
will have a bias towards programs closer to the upper bound  $\sampleMaxLength$. 
To mitigate this, we partition the interval $[\sampleMinLength, \sampleMaxLength)$ 
into equally sized \emph{buckets}, and we separately sample  in each bucket 
using \textsc{\algorithmOne{}}. By making the buckets sufficiently small, we can ensure 
that the overall sample will contain a balanced mix of programs of all sizes. 
Buckets will initially be consecutive in the index space, \ie, the upper bound
index one bucket will be the lower bound index of the next bucket, \etc. 
However, if the sampling algorithm cannot find enough programs on its first attempt,
we adjust bucket boundaries, and buckets may potentially overlap.
As a result, same program may be sampled from multiple buckets. 
However, as Algorithm \ref{algo:sample} filters out too short and too long programs, 
we won't have duplicates.
}

\subsection{Further Setup Details}
\label{subsec:further-setup}
To evaluate the validity of programs, we choose the following compilers. 
GCC for C, G++ for C++, OpenJDK for Java, .NET SDK for C\#, GFortran for Fortran, 
GnuCOBOL for COBOL, and GHC for Haskell. For Rust, Kotlin, Erlang, Swift, and Go, there is only
one standard compiler respectively, which we use. For all these compilers, we enable compiler    
flags to enforce the specific language standards. We conducted all experiments on 
a machine equipped with an AMD Ryzen Threadripper 2990 WX processor with 64 CPU 
cores and 128 GB of RAM running Ubuntu 22.04. We repeat all experiments 
\expReptitions{} times.  The relative standard deviations of CQ were at 
3.54\%, 3.24\%, and 2.46\% for Go, Swift, and Java respectively, and below 2\% 
for all other languages. We divide the sampled region of 0-256 bytes into 16 buckets of 16 bytes each.
Our target is to sample \sampleTarget{} programs for each bucket, with the exception
of Java and Kotlin, where due to performance issues we set a sample target 
of \sampleTargetJavaKotlin{} programs instead.

\section{Results}\label{sec:results}
\label{sec:result}
This section presents the results for 12 popular compiled languages 
based on the setup of the previous section. We first present CQs and then analyze LCQ plots to analyze 
the distribution of accepted programs for the different languages.

{\parindent0pt
\paragraph{\textbf{Result summary}}
\begin{itemize}
   \setlength\itemsep{0.58em}
   \item \emph{Large variety in CQ:} %
    C has a high CQ of \cCQ{}  followed by Erlang with \erlangCQ. By contrast, 
    Rust has the lowest CQ of \rustCQPrecise,  Swift has the second lowest CQ of 
    \swiftCQ. 
    \item \emph{Object-oriented languages have high CQ:} C\# (\csharpCQ),  
    Java (\javaCQ), C++ (\cppCQ), and Kotlin (\kotlinCQ) are listed in the  
    upper half of the ranking.  
    \item \emph{Distinct behavior of C:}  Strikingly, for C, the percentage of valid 
    programs stays high under increasing size. By contrast, for all other languages,  
    the LCQ for larger programs is near zero. 
\end{itemize}
}

\begin{figure}[t]
\begin{subfigure}{1.0\textwidth}
\[
\begin{array}{l@{\,=\,}r@{\qquad} l@{\,=\,}r@{\qquad} l@{\,=\,}r}
\cqSymbol{C} & \cCQPrecise & \qquad \cqSymbol{Erlang} & \erlangCQPrecise & \qquad \cqSymbol{C\#} & \csharpCQPrecise \\[0.25ex]
\cqSymbol{C++} & \cppCQPrecise & \qquad \cqSymbol{Kotlin} & \kotlinCQPrecise & \qquad \cqSymbol{Java} & \javaCQPrecise \\[0.25ex]
\cqSymbol{Haskell} & \haskellCQPrecise & \qquad \cqSymbol{Fortran} & \fortranCQPrecise & \qquad \cqSymbol{COBOL} & \cobolCQPrecise \\[0.25ex]
\cqSymbol{Go} & \goCQPrecise & \qquad \cqSymbol{Swift} & \swiftCQPrecise & \qquad \cqSymbol{Rust} & \rustCQPrecise \\[0.25ex]
\end{array}
\]
    \caption{}
    \label{fig:cq-values}
\end{subfigure}

    \vspace{0.5cm}
\begin{subfigure}{1.0\textwidth}
    \includegraphics[scale=0.57]{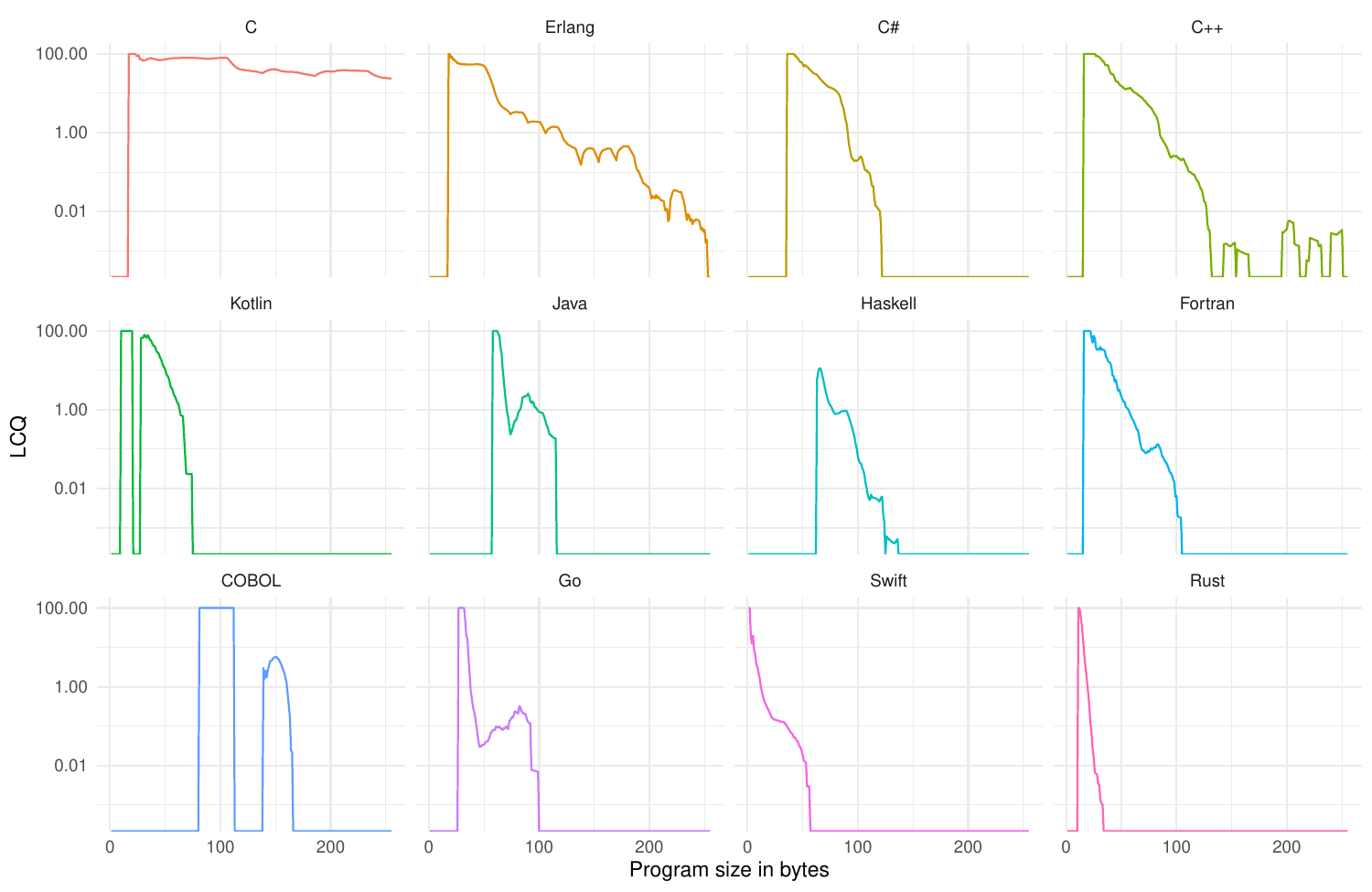}
    \caption{}
    \label{fig:cq-grid}
\end{subfigure}
    \caption{(a) Compilation quotients (CQs), and (b) average LCQ curves for the selected 
    programming languages.}   
    
\end{figure}

\begin{figure}[t!]
    \centering
    \includegraphics[width=0.8\textwidth]{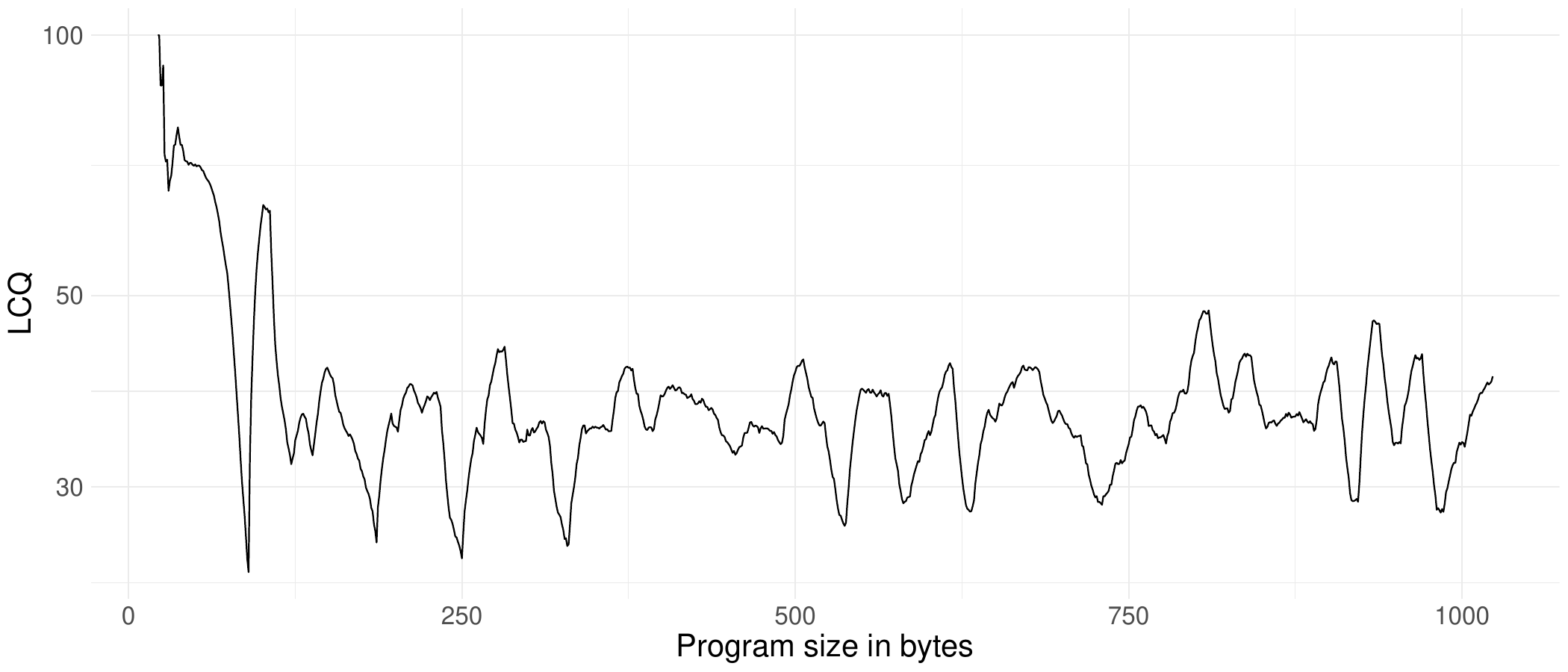}
    \caption{LCQ curve of C for a size up to \maxbytesC{} bytes}
    \label{fig:cq-c-large-limit}
\end{figure}

Fig~\ref{fig:cq-values} presents the CQ of each programming 
language. We observe that the CQ values are spread over a large range, with the 
majority of languages having a CQ below one. Only three languages are an exception to this: 
C, Erlang, and C++. The language with the lowest CQ is Rust at 0.0004. Strikingly, all 
object-oriented 
languages, \ie,  C\# (\csharpCQ),  Java (\javaCQ), C++ (\cppCQ), and Kotlin (\kotlinCQ) 
list in the upper half of the ranking. Moreover, newer languages such as Go (\goCQ), 
Swift (\swiftCQ) and Rust (\rustCQ) are listed at the bottom of the ranking.  
There are also languages commonly considered to be similar which exhibit
quite different CQ scores. For instance, C\#'s CQ is over five times that of Java, and 
C's CQ is over 80 times that of C++. To better understand the composition of the
CQ of each measured language, we now analyze their LCQ values. Fig.~\ref{fig:cq-grid} 
shows the LCQ as a function of size for each programming language. For almost 
every language, LCQ starts high \ie, near 100, and then declines to 
zero with some slope. A simple intuition explains this pattern: the longer the 
generated program, the more likely it is to contain an error.
An exception to this rule is C. Its LCQ, despite initially trending
downwards, does not reach zero. Instead, it stabilizes at a positive value. This
behavior continues beyond \maxbytes{} bytes and at least up to \maxbytesC{} bytes, 
as shown in Fig. \ref{fig:cq-c-large-limit}. Concretely, below 50 bytes, C's 
LCQ is above 70. Between 50 and 90 bytes, it decreases sharply to around 25,
followed by a sharp rise, and another sharp fall.
Subsequently, it varies between 25 and 50, without exhibiting an overall 
increasing or decreasing trend. This suggests that C's LCQ either 
never drops to zero, or does so only orders of magnitude slower than all 
other languages. 

Erlang lists second in the CQ ranking. Its LCQ curve behaves similarly to C's, 
however with important differences. Just like C's LCQ, it starts with a sharp drop,
but then rebounds and starts to oscillate. However, unlike in C's case, the drops 
are much sharper (\ie, the drop from around 50 to around 3 between 50 and 75 bytes), 
and the LCQ shows a clear negative trend as the size increases, reaching 
zero at approximately 253 bytes.
C and Erlang's oscillatory behavior already demonstrated that LCQ can be non-monotonic. 
These oscillations are unique to C and Erlang, but non-decreasing LCQ is not
-- many other tested languages exhibit non-decreasing regions \eg, Java, Fortran, and Go. COBOL's LCQ
curve also contains a significant non-decreasing region, yet it differs from
the other languages in that it is caused by a lack of programs in the search space, 
as described below. After initially hitting zero, the LCQ curve of most languages 
permanently remains 
there. 
C++'s LCQ is an outlier: after an initial drop to zero, it returns to a positive 
value and then again drops to zero. This pattern is repeated multiple times, with
the LCQ hovering on the order
of $10^{-3}$. This suggests that, in the relevant region, the true LCQ hovers 
close to the limit of what we can measure, leading to the apparent sudden 
jumps on the LCQ plot. COBOL's LCQ plot contains a region of zeroes between sizes 
of 113 and 138 bytes. Unlike for C++, this is not caused by all programs in the 
region being rejected, instead, there were no programs generated in this region. 
This is caused by COBOL's verbosity. Compared to other languages, COBOL's keywords 
are longer, leading to a less uniform distribution of program sizes.
Because of their similarity, we also briefly describe the differences in LCQ curves
of C\# and Java. Despite the difference in their CQ, their LCQ
curves have a similar structure. Both of them start from a near-100 value, drop
sharply, briefly stabilize, and then collapse to zero. However, Java's LCQ
drops quicker than C\#'s, leading to lower CQ.

\definecolor{codegreen}{rgb}{0,0.6,0}
\definecolor{codegray}{rgb}{0.5,0.5,0.5}
\definecolor{codepurple}{rgb}{0.58,0,0.82}
\definecolor{backcolour}{rgb}{0.95,0.95,0.92}

\lstdefinestyle{ccolor}{
    backgroundcolor=\color{white},
    commentstyle=\color{codegreen},
    keywordstyle=\color{blue},
    numberstyle=\tiny\color{codegray},
    stringstyle=\color{codepurple},
    basicstyle=\footnotesize\ttfamily,
    breakatwhitespace=false,
    breaklines=true,
    captionpos=b,
    keepspaces=true,
    numbers=left,
    numbersep=5pt,
    showspaces=false,
    showstringspaces=false,
    showtabs=false,
    tabsize=2
}

\lstdefinelanguage{Kotlin}{
  comment=[l]{//},
  commentstyle={\color{gray}\ttfamily},
  emph={filter, first, firstOrNull, forEach, lazy, map, mapNotNull, println},
  emphstyle={\color{OrangeRed}},
  identifierstyle=\color{black},
  keywords={!in, !is, abstract, actual, annotation, as, as?, break, by, catch, class, companion, const, constructor, continue, crossinline, data, delegate, do, dynamic, else, enum, expect, external, false, field, file, final, finally, for, fun, get, if, import, in, infix, init, inline, inner, interface, internal, is, lateinit, noinline, null, object, open, operator, out, override, package, param, private, property, protected, public, receiveris, reified, return, return@, sealed, set, setparam, super, suspend, tailrec, this, throw, true, try, typealias, typeof, val, var, vararg, when, where, while},
  keywordstyle={\color{NavyBlue}\bfseries},
  morecomment=[s]{/*}{*/},
  morestring=[b]",
  morestring=[s]{"""*}{*"""},
  ndkeywords={@Deprecated, @JvmField, @JvmName, @JvmOverloads, @JvmStatic, @JvmSynthetic, Array, Byte, Double, Float, Int, Integer, Iterable, Long, Runnable, Short, String, Any, Unit, Nothing},
  ndkeywordstyle={\color{BurntOrange}\bfseries},
  sensitive=true,
  stringstyle={\color{ForestGreen}\ttfamily},
}
\section{Analysis}\label{sec:posthoc}
In this section, we interpret the results from Section~\ref{sec:results}.
We first analyze the two extremes, C and Rust, and then move on to the remaining languages,
comparing the 
behavior of related languages throughout, and finally explain CQ-affecting features.

\subsection{C and Rust -- the extremes}
As we observed, not only is C's CQ much higher than that of all other languages, 
but its LCQ curve is also peculiar: unlike all other languages, it does not approach 
zero. On the other extreme, we have Rust, for which only 6 out of 1.5 
million programs compile. We begin our analysis with C 
(see Fig. \ref{fig:c-examples}). To better understand C's behavior, 
we first look at programs below 50 bytes consisting of only one or 
two statements. Our first example (Fig~\ref{prog:c-goodsmalldecl}) 
shows a valid program with a declaration of a short pointer \texttt{\small var1},
and the second is an invalid program with a single continue statement appearing outside 
of a loop (Fig.~\ref{prog:c-continue}). The third one is an invalid program 
(Fig.~\ref{prog:c-defuse}) with a goto statement to label \texttt{\small var1}. 
Since \texttt{\small var1} is undeclared, the program is invalid. Another, more 
complex example is depicted in Fig.~\ref{prog:c-badsmalldecl}. The program
defines an integer pointer with the \texttt{\small \_Thread\_local} modifier, which 
indicates there will be a separate version of the pointer for each  
thread \texttt{\small main} spawns. However, the program is invalid since 
\texttt{\small \_Thread\_local} variables must be either \texttt{\small static} or
\texttt{\small extern}. Looking at larger programs, we observe that declarations 
dominate C programs, outnumbering all other types of statements.
One reason for this is that C's declarations are flexible, consist 
of many optional elements (specifiers), and can be arbitrarily nested.
Figs. \ref{prog:c-accepted-decl}-\ref{prog:c-badtype} demonstrate this.   
Fig.~\ref{prog:c-accepted-decl} presents a valid declaration, nesting several pointer
declarations. Fig. \ref{prog:c-accepted-typedef} likewise presents a valid 
program, which declares a \texttt{\small typedef} instead of a variable.
Fig. \ref{prog:c-badrestrict} presents an invalid program which 
is invalid due to incorrect use of the \texttt{\small restrict} keyword
which is not allowed to be applied directly to a non-pointer type
(in this case, \texttt{\small unsigned}). The example in Fig.
\ref{prog:c-badtype} is invalid because 
\texttt{\small var2} is not a type. Fig.~\ref{fig:c-examples-large}
illustrates a deeply nested but valid pointer declaration.

\newcommand{\firstcolspacer}{-0.4cm}
\newcommand{\midcolspacer}{-0.6cm}
\newcommand{\spacer}{0.1cm}

\begin{figure}
\begin{subfigure}{0.25\textwidth}
\centering
\begin{minted}{C}
/* valid */ 
int main(void) {
    register short *var1;
}
\end{minted}
\vspace{\firstcolspacer}
\caption{}
\label{prog:c-goodsmalldecl}
\end{subfigure}
\hfill
\begin{subfigure}{0.2\textwidth}
\centering
\begin{minted}{C}
/* invalid */ 
int main(void) {
    continue;
}
\end{minted}
\vspace{\firstcolspacer}
\caption{}
\label{prog:c-continue}
\end{subfigure}
\hfill
\begin{subfigure}{0.2\textwidth}
\centering
\begin{minted}{C}
/* invalid */
int main(void) {
    goto var1;
}
\end{minted}
\vspace{\firstcolspacer}
\caption{}
\label{prog:c-defuse}
\end{subfigure}
\vspace{\spacer}

\begin{subfigure}{0.7\textwidth}
\centering
\begin{minted}{C}
              /* invalid */
              int main(void) {
                  _Thread_local restrict int *var2;
              }
\end{minted}
\vspace{\firstcolspacer}
\caption{}
\label{prog:c-badsmalldecl}
\end{subfigure}

\begin{subfigure}{1.0\textwidth}
\centering
\begin{minted}{C}
/* valid */
int main(void) {
    extern const _Atomic long *const _Atomic *volatile _Atomic *const volatile _Atomic *const restrict volatile var1;
}
\end{minted}
\vspace{\midcolspacer}
\caption{}
\label{prog:c-accepted-decl}
\end{subfigure}
\vspace{\spacer}

\begin{subfigure}{1.0\textwidth}
\begin{minted}{C}
/* valid */
int main(void) {
    typedef const volatile unsigned *const restrict volatile _Atomic *const volatile _Atomic *restrict _Atomic *volatile _Atomic var1;
}
\end{minted}
\vspace{\midcolspacer}
\caption{}
\label{prog:c-accepted-typedef}
\end{subfigure}
\vspace{\spacer}

\begin{subfigure}{1.0\textwidth}
\begin{minted}{C}
/* invalid */
int main(void) {
    static restrict volatile _Atomic unsigned **restrict volatile _Atomic *volatile *const restrict volatile var1;
}
\end{minted}
\vspace{\midcolspacer}
\caption{}
\label{prog:c-badrestrict}
\end{subfigure}
\vspace{\spacer}

\begin{subfigure}{1.0\textwidth}
\begin{minted}{C}
/* invalid */
int main(void) {
    extern volatile _Atomic var2 *const restrict *restrict volatile *volatile _Atomic *const volatile _Atomic var1;
}
\end{minted}
\vspace{\midcolspacer}
\caption{}
\label{prog:c-badtype}
\end{subfigure}

\caption{Assorted C programs generated by \toolname{}.}
\label{fig:c-examples}
\end{figure}

\begin{figure}
    \begin{minted}{C}
/* valid */
int main(void) {
    extern volatile _Complex *const volatile _Atomic *const restrict volatile _Atomic *const *const volatile _Atomic *const restrict *const restrict _Atomic *restrict volatile _Atomic *const restrict volatile _Atomic *volatile *const restrict _Atomic *const volatile _Atomic *const restrict _Atomic *const restrict _Atomic *const restrict _Atomic *const _Atomic(*volatile _Atomic *const *volatile _Atomic **const _Atomic *_Atomic *const restrict *const restrict volatile _Atomic *const restrict *restrict volatile **const restrict volatile _Atomic *restrict volatile _Atomic *const restrict volatile _Atomic *restrict _Atomic *const **restrict volatile _Atomic *const restrict *volatile *_Atomic *const restrict _Atomic *const *const volatile(*const restrict var2));
}
\end{minted}
    \vspace{-0.3cm}
    \caption{A 780-byte long, valid C program.}
    \label{fig:c-examples-large}
 \end{figure}

Another key aspect of C's declarations is that they impose  
few conditions. Particularly for pointer declarations, which dominate longer
programs, the number of conditions does not increase as they are nested. 
To understand this, let us contrast this with nested arithmetic expressions for which the 
constraints do increase. Consider an expression \texttt{\small a + b}. It is valid 
if (1) \texttt{\small a} and \texttt{\small b} are individually valid, 
and (2) \texttt{\small a} and \texttt{\small b} have types that can be added together. 
Condition (2) is imposed on the subexpressions by its parent.
If these subexpressions are themselves composed of other expressions, they may 
have to impose additional conditions on their subexpressions
to satisfy condition (2). Hence the number of conditions may grow as the expression
becomes longer, making it harder to generate valid nested expressions.
By contrast, in a C pointer declaration \texttt{\small T *D}, no conditions 
are imposed on declarator \texttt{\small D}. Hence, compiling nested 
pointer declarations does not become harder as declarations get longer. 

\begin{figure}[t!]
    \centering
    \includegraphics[width=0.8\textwidth]{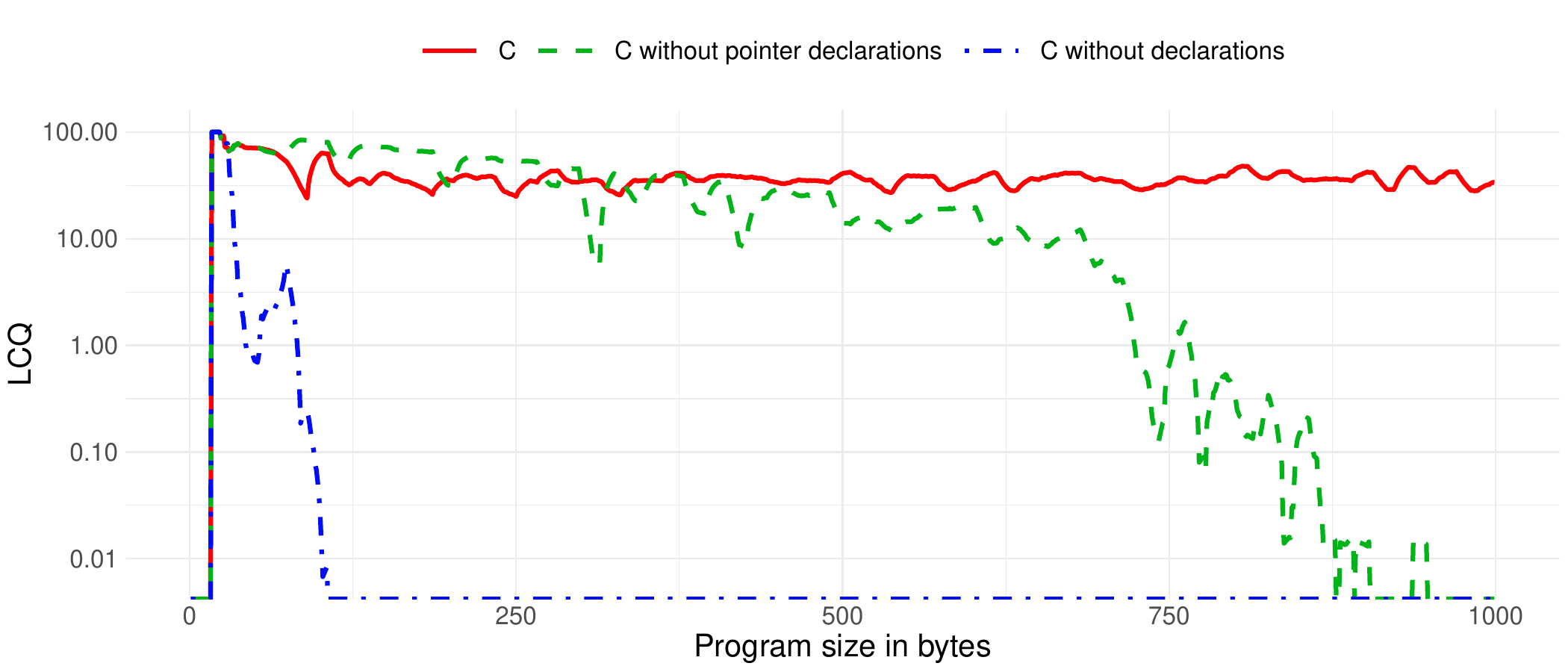}
    \caption{LCQ curve of C, C without pointer declarations, and C without declarations, up to \maxbytesC{}
    bytes}
    \label{fig:c-variants}
\end{figure}

To examine whether this behavior leads indeed to C's higher LCQ, we calculated 
the LCQ of two variants of C: one without pointer declarations, and one without 
any declarations. Fig. \ref{fig:c-variants} presents the LCQ of these variants up 
to \maxbytesC{} bytes. Without any declarations, C's LCQ falls quickly to zero. The variant without
pointer declarations (but with all other types of declarations) exhibits a significantly
higher LCQ, but it also plummets to zero at around 950 bytes. This shows that pointer 
declarations are essential for C's LCQ not to plummet.
In many ways, Rust behaves differently from C. The generated Rust programs are dominated 
by operator expressions, which, as explained above, are less likely valid under increasing length.  Furthermore, Rust's type system is strict; it does not accept operands 
of mismatching types. For example, \texttt{\small 1u32 + 2u8} and \texttt{\small 
1u32 + ('a' == 'a')} is invalid Rust, but valid C. This means that, while most of the programs are operator expressions, the probability that such expressions compile is extremely small. This in turn leads to very low CQ. In essence, both C and Rust have a construction that dominates
as program size increases, but C's programs are likely valid, while Rust's
programs become unlikely to be valid with growing size. 

Fig. \ref{fig:rust-examples} presents Rust examples. Rust's type system permits
some constructions that would not be valid in most other imperative languages. For 
example, Fig. \ref{prog:rust-returnnot} shows a valid program that negates the result of 
a return expression. \texttt{\small return} has type \texttt{\small !} (never/bottom), which
implements the \texttt{\small Not} trait, overloading the negation operator. Thus
\texttt{\small !return} has type \texttt{\small !}, which is coercible to
\texttt{\small ()}. \texttt{\small main}'s return type is \texttt{\small ()}, thus the
program is valid.
Fig. \ref{prog:rust-returncast} presents a similar example. \texttt{\small return as \_}
casts \texttt{\small return} to an inferred type. This type is inferred to be
\texttt{\small ()}, as that \texttt{\small main}'s return type. Since \texttt{\small !}
is castable to any type, the program is valid. The majority of Rust samples, however,
are invalid. Fig. \ref{prog:rust-badadd} shows an
invalid operator expression, adding a 32-bit integer to a 16-bit one. 
Fig. \ref{prog:rust-badexp} presents a longer example, which divides a string by an integer
and assigns to a temporary value, and is thus invalid.

\begin{figure}
    \begin{subfigure}{0.2\textwidth}
        \begin{minted}{Rust}
// valid  
fn main() { 
    !return  
} 
        \end{minted}
         \vspace{-0.2cm}
        \caption{}
        \label{prog:rust-returnnot}
    \end{subfigure}
    \hfill
    \begin{subfigure}{0.2\textwidth}
        \begin{minted}{Rust}
// valid 
fn main() { 
    return as _
}
        \end{minted}
         \vspace{-0.2cm}
        \caption{}
        \label{prog:rust-returncast}
    \end{subfigure}
    \hfill
    \begin{subfigure}{0.2\textwidth}
        \begin{minted}{Rust}
// invalid  
fn main() {
    42u32 + 42u16
}
        \end{minted}
         \vspace{-0.2cm}
        \caption{}
        \label{prog:rust-badadd}
    \end{subfigure}

    \vspace{\spacer}
    \begin{subfigure}{0.78\textwidth}
        \begin{minted}{Rust}
                // invalid 
                fn main() { 
                    "string" / 42isize >= 42i128 ^= 123.4f32 /= 42u16
                }
        \end{minted}
         \vspace{-0.2cm}
        \caption{}
        \label{prog:rust-badexp}
    \end{subfigure}
    \caption{Assorted Rust programs found by \toolname{}.}
    \label{fig:rust-examples}
\end{figure}

\subsection{Comparison of related languages}
This section compares the behavior of related languages such as C and C++, 
Java and C\#, C++ and C\#. Fortran, COBOL, Go and Swift are treated 
in Section \ref{section:appendix}. 

{\parindent0pt 
\paragraph{\textbf{C and C++}} C++ is almost a superset of C. 
Hence, it is striking that C++'s CQ is so much smaller than that of C. 
Furthermore, C++ also has the pointer declarations causing C's non-collapsing LCQ curve. 
It is thus natural to ask: \emph{Why does C++'s LCQ curve plummet to zero while C's curve does not?} 
As our results reveal, the answer is that C++ has other features 
that dominate pointer declarations, but are unlikely to be valid. These features 
include namespace definitions and using declarations. Furthermore, within these 
constructions, other flexible, but hard to compile features are nested:   
qualified identifiers, \eg, \texttt{\small var2::var1::var1::var2::var1}, templates,
and operator function identifiers, \eg, \texttt{\small operator +} occur frequently.
Programs including these features are less likely to compile than C's pointer declarations. 
In particular, qualified identifiers are highly unlikely to compile, 
as doing so requires multiple matching namespace declarations. 
}
Fig.~\ref{prog:cpp-twostatements} presents an example which first declares a label  
\texttt{\small var1}, then uses a using-declaration bringing \texttt{\small ::operator new[]}
into the current scope (this is essentially a no-op, as the operator is already in the global scope).
It then also declares an empty \texttt{\small enum class}.
Fig.~\ref{prog:cpp-manyusing} also depicts a valid program containing 
three using-declarations, the first one with a label.  Fig. \ref{prog:cpp-qualified-ident} 
presents an example with a generic operator, with the generic parameter itself 
being an operator nested within several namespaces. Neither the referenced 
namespaces nor the operators exist, thus the program is invalid.
Fig. \ref{prog:cpp-badfriend} presents a declaration that uses many modifiers, and is invalid
due to the use of the \texttt{\small friend} modifier outside a class. 
Despite the complications with matching namespaces \etc, C++ still has a
high CQ compared to other languages, as smaller programs do compile 
frequently. 
These programs contain empty blocks (\texttt{\small\{\}}), 
simple declarations, \etc  similar to C.

\begin{figure}
\begin{subfigure}{0.35\textwidth}
\begin{minted}{C++}
// valid 
int main() {
    var1: using ::operator new[];
    enum class var1;
}
\end{minted}
\vspace{-0.2cm}
\caption{}
\label{prog:cpp-twostatements}
\end{subfigure}
\hspace{0.25cm}
\begin{subfigure}{0.45\textwidth}
\centering
\begin{minted}{C++}
// valid 
int main() {
    var2: var1: using ::operator delete[];
    using ::operator delete;
    using ::operator delete[];
}
\end{minted}
\vspace{-0.2cm}
\caption{}
\label{prog:cpp-manyusing}
\end{subfigure}

\vspace{0.2cm}
\begin{subfigure}[b]{1.0\textwidth}
\begin{minted}{C++}
// invalid 
int main() {
    using ::operator!<::var1::var2<>::var1::var1::var2::template operator^=...>;
}
\end{minted}
\vspace{-0.4cm}
\caption{}
\label{prog:cpp-qualified-ident}
\end{subfigure}

\vspace{0.2cm}
\begin{subfigure}[b]{1.0\textwidth}
\begin{minted}{C++}
// invalid 
int main() {
    friend register static thread_local extern mutable inline virtual explicit typename var1::var2::template var2<>;
}
\end{minted}
\vspace{-0.6cm}
\caption{}
\label{prog:cpp-badfriend}
\end{subfigure}
\label{fig:cpp-examples}
\vspace{-0.4cm}
\caption{Assorted C++ programs found by \toolname{}.}
\end{figure}

{\parindent0pt 
\paragraph{\textbf{C\# and Java}} C\# and Java are both popular object-oriented languages, with many
similarities. Despite this, C\# has higher CQ than Java, but their LCQ curve
is quite similar.   
C\#'s higher CQ can be explained by its LCQ curve decreasing less rapidly. One
factor contributing to this smaller slope is that C\#'s entry point is shorter than
Java's. This does not necessarily imply a higher CQ, but C\#'s small programs are
also likely to compile, whereas in Java's case, we can observe an early drop in LCQ. 
Fig. \ref{fig:csharp-examples} and \ref{fig:java-examples} present C\# and 
Java programs respectively. At smaller sizes, C\# programs 
consist of many declarations, similar to C and C++. In particular, multi-dimensional array
declarations, such as in Fig \ref{prog:cs-multiarr}, are common. 
By contrast, Java programs consist of expressions or control-flow statements that 
contain expressions, as seen in Fig. \ref{prog:java-goodassert}. Although Java's
type system is lax for expressions \eg permitting \texttt{\small 'a' == 123.4d},
many expressions are still invalid due to type errors. Consider, \eg,  
Fig.~\ref{prog:java-badifthrow}, an invalid program with an if-statement that   
has a double in its condition. 
This trend extends to larger programs with longer expressions which are 
more likely to contain errors. An example is shown in Fig. \ref{prog:java-badexp}, which contains
an invalid use of the increment operator, and several instances of incorrect operand
types. A longer valid Java program is depicted in Fig. \ref{prog:java-goodloop},
with correct usage of expressions and looping.
In C\#'s case, longer programs are dominated by generic local 
function declarations instead of expressions. These can be made arbitrarily long 
by adding more generic constraints, which are unlikely to compile. Fig. \ref{prog:cs-goodgeneric} and \ref{prog:cs-badgeneric} present a valid and invalid example of generic local functions respectively. 
Fig~\ref{prog:cs-goodgeneric} shows a valid C\# program 
declaring a local function \texttt{\small var1}, with a generic parameter also named
\texttt{\small var1}, which is bound by two constraints: 
\texttt{\small var1: class?}, whic  prescribes that it must be a (potentially nullable) 
reference type, and \texttt{\small var1: var1}, which prescribes that it must be 
\texttt{\small var1} or a subclass of \texttt{\small var1}.
Fig.~\ref{prog:cs-badgeneric} declares a local function, but it attempts to constrain nonexistent
generic parameters, and thus is invalid. The latter type of programs is much more common than the
former, hence C\#'s LCQ plummets to zero at large program sizes.
}

\begin{figure}
\begin{subfigure}{0.35\textwidth}
        \begin{minted}{csharp}
// valid
void Main() {
    ulong [,,,,] var2;
}
\end{minted}
\vspace{-0.4cm}
\caption{}
\label{prog:cs-multiarr}
\end{subfigure}
\hfill
\begin{subfigure}{0.55\textwidth}
\begin{minted}{csharp}
// valid 
void Main() {
    void var1<var1>() where var1 : class?, new() {}
}
\end{minted}
\vspace{-0.4cm}
\caption{}
\label{prog:cs-goodgeneric}
\end{subfigure}
\vspace{0.4cm}

\begin{subfigure}[b]{0.65\textwidth}

    \begin{minted}{csharp}
// invalid
void Main() {
    void var1() where var2 : new() where var1 : class?, var1 {}
}
\end{minted}
\vspace{-0.4cm}
\caption{}
    \label{prog:cs-badgeneric}
\end{subfigure}

\caption{Assorted C\# programs found by \toolname. The class enclosing the \texttt{\small Main} functions is omitted for brevity.}
\label{fig:csharp-examples}
\end{figure}

\begin{figure}
\begin{subfigure}{0.28\textwidth}
\begin{minted}{Java}
// valid
void main(String[] args) {
   assert 'a' == 123.4d;
}
\end{minted}
\vspace{-0.4cm}
\caption{}
\label{prog:java-goodassert}
\end{subfigure}
\hfill
\begin{subfigure}{0.3\textwidth}
\begin{minted}{Java}
// invalid
void main(String[] args) {
  if(123.4) 
    throw true;
}
\end{minted}
\vspace{-0.4cm}
\caption{}
\label{prog:java-badifthrow}
\end{subfigure}
\hfill
\begin{subfigure}{0.38\textwidth}
\begin{minted}{Java}
// valid 
void main(String[] args) {
  while(-'a' <= + 123.4d + 'a') var2: 
    continue;
}
\end{minted}
\vspace{-0.4cm}
\caption{}
\label{prog:java-goodloop}
\end{subfigure}

\vspace{0.2cm}
\begin{subfigure}{1.0\textwidth}
\begin{minted}{Java}
// invalid
void main(String[] args) {
    assert ++123.4d % true && 123.4d - 123.4d >= 42 / 'a' || 123.4d > 123.4f << 123.4d;
}
\end{minted}
\vspace{-0.4cm}

\caption{}
\label{prog:java-badexp}
\end{subfigure}

\caption{Assorted Java programs found by \toolname. The class enclosing the \texttt{\small main} functions is omitted for brevity.}
\label{fig:java-examples}
\end{figure}

\begin{figure}
\begin{subfigure}{0.3\textwidth}
\begin{minted}{Kotlin}
// valid 
fun main() {
  open class var1 <out var1>
}
\end{minted}
\caption{}
\label{prog:kotlin-simple}
\end{subfigure}
\hfill
\begin{subfigure}{0.28\textwidth}
\begin{minted}{Kotlin}
// invalid  
fun main() {
  interface var2
}
\end{minted}
\vspace{-0.4cm}
    \caption{}
    \label{prog:kotlin-local-interface}
\end{subfigure}
\begin{subfigure}{0.4\textwidth}
    \begin{minted}{Kotlin}
// valid 
fun main() {
  var1@ var2@ var2@ var1@ class var2
}
\end{minted}
\caption{}
\label{prog:kotlin-labels}
\end{subfigure}

\begin{subfigure}[b]{1.0\textwidth}
\begin{minted}{Kotlin}
// invalid 
fun main() {
  open inner internal abstract class var2<in *, out var1,> override operator infix inline suspend abstract final constructor(var var2: dynamic)
}
\end{minted}
\vspace{-0.4cm}
\caption{}
\label{prog:kotlin-long-invalid}
\end{subfigure}

\caption{Assorted Kotlin programs generated by \toolname.}
\label{fig:kotlin-examples}
\end{figure}

{\parindent0pt
\paragraph{\textbf{C\# and C++}}
C\# and C++ are close in the CQ ranking. Indeed, C\#'s CQ
is closer to C++'s than Java's. This is because both languages are prone to generating
declarations instead of expressions. Declarations are more likely to compile, 
as they have few conditions imposed on them. Interestingly, while C++ programs often
contain  qualified identifiers, these are seldom generated in C\#.
}

{\parindent0pt
\paragraph{\textbf{Kotlin and Java}}
Kotlin combines object-oriented paradigms with functional ones 
and compiles to JVM. Compared to Java, it has a slightly higher CQ. While Java's 
programs are dominated by expressions, Kotlin's programs are instead dominated 
by declarations. This is explained by the fact that Kotlin has many more 
types of declarations and modifiers than Java, giving the generator more options.
Fig. \ref{prog:kotlin-simple} presents a simple valid declaration, while Fig. 
\ref{prog:kotlin-local-interface} presents a simple declaration that is invalid, as 
interfaces cannot be defined locally within a function. Fig. \ref{prog:kotlin-long-invalid}
is an example of a long, invalid declaration. The program attempts to define a 
local generic class with a constructor.  However, it has many errors, 
including the usage of modifiers that are not applicable, incorrect use of star 
projections, and incorrect use of dynamic types. A feature that increases 
Kotlin's CQ, however, is that multiple labels are allowed to have the
same name, unlike in Java. Fig. \ref{prog:kotlin-labels} presents an example that
reuses two label names on a declaration. Overall, because of its richer declarations, Kotlin
behaves more similarly to C or C\# than Java. However, Kotlin's declarations are less likely to compile
than those of languages with higher CQs.
}

\begin{figure}
\begin{subfigure}{0.25\textwidth}
\begin{minted}{Erlang}
% valid
func() -> + + + + - 42.
\end{minted}
\caption{}
\label{prog:erl-goodunary}
\end{subfigure}
\hfill
\begin{subfigure}{0.7\textwidth}
\begin{minted}{Erlang}
% valid
func() -> catch catch not not not bnot fun an_atom:an_atom/42.
\end{minted}
\caption{}
\label{prog:erl-goodcatchfun}
\end{subfigure}

    \vspace{0.4cm}
\begin{subfigure}{0.3\textwidth}
        \begin{minted}{Erlang}
% invalid
func() -> + + + - + - - Var2.
\end{minted}
\caption{}
\label{prog:erl-defuse}
\end{subfigure}
\hfill
\begin{subfigure}{0.69\textwidth}
    \begin{minted}{Erlang}
% invalid
func() -> catch bnot + fun an_atom:an_atom/42:"string" "string".
\end{minted}
    \caption{}
    \label{prog:erl-badremotecall}
\end{subfigure}

\caption{Assorted Erlang programs generated by \toolname.}
\label{fig:erlang-examples}
\end{figure}
\vspace{0.2cm}


\begin{figure}
    \begin{subfigure}{0.3\textwidth}
        \begin{minted}{Haskell}
-- valid
func = func
    where { func = -func; }
\end{minted}
        \caption{}
        \label{prog:hs-numrec}
    \end{subfigure}
    \begin{subfigure}{0.3\textwidth}
        \begin{minted}{Haskell}
-- invalid
func = 42
    where { var1 = -'a'; }
\end{minted}
        \caption{}
        \label{prog:hs-typeerror}
    \end{subfigure}
    \vspace{0.2cm}

    \begin{subfigure}{0.49\textwidth}
        \begin{minted}{Haskell}
-- valid
func = var2 
    where { var2 = (#,#) (,,,,,) {}; }
\end{minted}
        \caption{}
        \label{prog:hs-recordupdate}
    \end{subfigure}
    \begin{subfigure}{0.49\textwidth}
        \begin{minted}{Haskell}
-- invalid
func = 42
    where { var2 = mdo {} @ 'var1; }
\end{minted}
        \caption{}
        \label{prog:hs-template}
    \end{subfigure}

    \begin{subfigure}{1.0\textwidth}
        \begin{minted}{Haskell}
-- invalid
func = var2
    where { func = func :: forall. forall -> var2 :: - '`func` + `func` : + `var1` + : '`func` '`func` - `var1` ' + -; }   
\end{minted}
        \vspace{-0.2cm}
        \caption{}
        \label{prog:hs-templatetyapps}
    \end{subfigure}

\caption{Assorted Haskell programs generated by \toolname.}
\label{fig:haskell-examples}
\end{figure}

{\parindent0pt
\paragraph{\textbf{Erlang and Haskell}} Erlang and Haskell are purely 
functional languages. Their CQs are very different, with Erlang having the second highest CQ
among tested languages, and Haskell having the sixth lowest. Erlang's high  
CQ is mostly a result of its dynamic typing, because of which even clearly type-incorrect 
programs
are deemed valid (but will raise type errors on execution). Erlang programs necessarily
consist only of expressions, as nothing else is permitted inside a function. As a result,
Erlang exhibits an initially high LCQ, which falls slowly. Fig.~\ref{prog:erl-goodunary}
presents 
a valid program which defines a function returning \texttt{\small -42}. The multiple 
plus operators have no effect. Fig.~\ref{prog:erl-goodcatchfun} presents a program 
that is deemed valid  but will crash at runtime because of type errors. It also 
creates an anonymous function that calls the function \texttt{\small an\_atom:an\_atom} 
with arity 42. This function does not exist, which is another reason why the program 
fails at runtime.
Fig.~\ref{prog:erl-defuse} shows an example similar to Fig. \ref{prog:erl-goodunary}, except
it is invalid due to its use of an undeclared identifier.
Fig.~\ref{prog:erl-badremotecall} depicts another, more elaborate 
invalid program. The code attempts to perform a function call of the form
\texttt{\small <Module>:<Function>}. However, it specifies a   
function instead of a module name and a string instead of a function identifier.
This is syntactically valid as \texttt{\small Module} and \texttt{\small Function} can 
arbitrary expressions. Furthermore, the expression \texttt{\small "string" "string"} is 
valid and equivalent to \texttt{\small "stringstring"}.
}

Different from Erlang, Haskell is statically typed. 
Haskell's LCQ steeply decreases as more and more programs 
encounter type errors. However, recursion and Haskell's flexible type system allow some
programs to compile, leading to a higher CQ than other languages. 
Fig. \ref{prog:hs-numrec} presents a Haskell program which recurses 
to make \texttt{\small func} a value of type \texttt{\small Num a => a}.
The program is valid but does not terminate. Fig. \ref{prog:hs-typeerror} 
depicts a program which is invalid because of a type error, as it attempts to negate the 
character \texttt{\small 'a'}. Fig. \ref{prog:hs-template} is a more complicated 
invalid program: it contains an empty \texttt{\small mdo} block, and a 
template Haskell quotation of \texttt{\small var1}, which is undefined. 
Tuples are also often used in longer Haskell programs. 
Fig. \ref{prog:hs-recordupdate} presents a valid 
program combining several elements -- \texttt{\small (,{},{},{},{},) \{\}}
performs a "construction using field labels" \cite[Section~3.15.2]{marlow2010haskell}
of a 6-tuple without specifying any field labels, resulting in a tuple of 
type \texttt{\small (a,b,c,d,e,f)}. This tuple is then applied to the unboxed 
binary tuple constructor \texttt{\small (\#,\#)}, resulting in a type 
of \texttt{\small g -> (\# (a,b,c,d,e,f), g \#)}. The function would raise 
an error on execution, as the constructor \texttt{\small (,{},{},{},{},)}
is not a record constructor, and therefore the elements of the tuple are not properly
initialized. Finally, programs with Template Haskell quotes, backticks,
and type applications are rejected by GHC (see Fig. \ref{prog:hs-templatetyapps}).

\subsection{Analysis of Fortran, Cobol, Go and Swift}\label{section:appendix}
We consolidate Fortran, COBOL, Go and Swift. Fortran
programs are dominated by type declarations, which have many features and parameters, and thus
seldom compile. Fig. 
\ref{prog:fortran-goodtype} is a valid example, declaring a multidimensional character
array, while Fig. \ref{prog:fortran-badtype} presents an example that is invalid because it
uses a complex literal and a string as an array length.
COBOL's CQ is extremely close to that of Fortran, and its behavior is also similar, with 
declarations dominating. In particular, \texttt{\small RENAMES} clauses are common,
and are prone to errors due to undeclared identifiers and other semantic errors (\eg Fig.
\ref{prog:cobol-renames}).
Fig. \ref{prog:cobol-gooddecl} presents a valid program that declares a national
data item. 

\begin{figure}[h!]
\begin{subfigure}{0.45\textwidth}
\begin{minted}{Fortran}
! Fortran, valid
PROGRAM MAIN
TYPE var1
CHARACTER (LEN = *), POINTER :: var1(:) * 42
ENDTYPE
END
\end{minted}
\caption{}
\label{prog:fortran-goodtype}
\end{subfigure}
\hfill
\begin{subfigure}{0.53\textwidth}
\begin{minted}{Fortran}
! Fortran, invalid
PROGRAM MAIN
TYPE var1
CHARACTER * (-42, -42) :: var1(:) * 'string'
ENDTYPE
END
\end{minted}
\caption{}
\label{prog:fortran-badtype}
\end{subfigure}

\vspace{0.2cm}
\begin{subfigure}{0.48\textwidth}
\begin{minted}{Cobolfree}
* COBOL, valid
DATA DIVISION.
WORKING-STORAGE SECTION.  
    77 var2 PICTURE N.
PROCEDURE DIVISION.
END PROGRAM test.
\end{minted}
\caption{}
\label{prog:cobol-gooddecl}
\end{subfigure}
\hfill
\begin{subfigure}{0.48\textwidth}
\begin{minted}{Cobolfree}
* COBOL, invalid
DATA DIVISION. 
WORKING-STORAGE SECTION.
66 var1 RENAMES var1 OF var1 THROUGH var1 OF var2.  
66 var2 RENAMES 42 IN 42 THROUGH 42 OF var2. 
PROCEDURE DIVISION.
END PROGRAM test.
\end{minted}
\caption{}
\label{prog:cobol-renames}
\end{subfigure}

    \begin{subfigure}{1.0\textwidth}
        \begin{minted}{Go}
// Go, invalid
func main() {
    go + & & var1.var2.var1 &^ <- & + * - <- ^ * ! + ! ^ & <- var1.var2.var1 && * ! + - - <- * ^ ^ * nil.var1 >> & "string".var2
}
        \end{minted}
        \vspace{-0.4cm}
        \caption{}
        \label{prog:go-longexpr}
    \end{subfigure}

    \vspace{0.2cm}
\begin{subfigure}{0.2\textwidth}
        \begin{minted}{Swift}
// Swift, valid
do {} catch {};
        \end{minted}
        \caption{}
        \label{prog:swift-docatch}
\end{subfigure}
\hfill
\begin{subfigure}{0.5\textwidth}
        \begin{minted}{Swift}
    // Swift, invalid
    try var1(var2:).
        init.init($1:).
            init($0:).
                init($0:).init(var2:);
        \end{minted}
        \caption{}
        \label{prog:swift-init}
\end{subfigure}
\begin{subfigure}{0.25\textwidth}
        \begin{minted}{Go}
// Go, valid
func main() {
    goto var1
var1:
    select {}
    goto var1
}
        \end{minted}
        \caption{}
        \label{prog:go-select}
\end{subfigure}
\caption{Assorted Fortran, COBOL, Go, and Swift programs generated by \toolname.}
\label{fig:other-examples}
\end{figure}

Unlike the other two languages, Go is dominated by expressions and some control-flow 
structures. Srict typing and features that require a particular type of expression lead
to a low CQ. Fig. \ref{prog:go-select} 
presents a valid program which correctly uses \texttt{\small goto} statements. Fig.
\ref{prog:go-longexpr} presents an expression that is invalid because (among other errors)
the argument of its \texttt{\small go} statement is not a function call.
Finally, Swift achieves the second-lowest CQ of all languages due to frequent use of
initializer prefixes, as well as the use of implicit parameter names
(\eg \texttt{\small \$0}) outside closures. Fig. \ref{prog:swift-docatch} is a simple
valid example, while Fig. \ref{prog:swift-init} is an invalid example that demonstrates
the issues mentioned above.

\subsection{Analysis of CQ-affecting language features}
A common observation in high-CQ languages are 
(1) flexible features, \ie features that can grow arbitrarily, and (2) easy-to-compile
features, \ie a random
instance of the given feature is likely accepted by the compiler. In C's case, these are pointer 
declarations, in Erlang's case these are expressions. 
Conversely, low-CQ languages have features that are flexible, 
but hard to compile.  For Rust and Swift, these are expressions. 
The languages with less extreme CQs have mixes of different
features. For instance, a language can start with 
easy-to-compile features at small sizes, but at higher sizes new structures 
become available that are more flexible and harder to compile (\eg C\#). 
The two most influential features for CQ are expressions and declarations.
Expressions are flexible since they can be nested in all languages \eg operator expressions
and function calls. Their compilation hardness is determined by the strictness of the type
system.  Declarations, on the other hand, vary in their flexibility. 
Their compilation hardness is usually not caused by the type system. 
Other common features, such as control-flow statements are less frequent.

Languages with more complex features usually also allow more complex declarations. 
This leads to more restrictions, thus lower CQ. Another important aspect is whether
declarations 
are allowed inside functions. For example, Rust has no shortage of 
complex features (\eg, traits), yet they cannot be declared within functions. 
Thus, they also do not affect the CQ of the language. Other languages, such as Swift, 
permit a high variety of declarations within functions, increasing the importance 
of declarations to CQ. In a nutshell, highly general constructs cause lower CQ. 
On the other hand, the strictness of the type system is secondary, 
mostly influencing how difficult expressions are to compile. 
For C, the modest complexity of constructs leads to easy-to-compile 
declarations, while for Erlang, the lack of static typing leads to easy-to-compile 
expressions, both leading to high CQ.

\section{Discussion}\label{sec:discussion}
Having analyzed CQ values and LCQ graphs of programming languages, we discuss 
CQ's implications on (1) programmers, (2) the long-run adoption of programming languages,     
and, (3) the fuzz-testing of compilers. Finally, we discuss the limitations of our     
work.

{\parindent0pt 
\paragraph{\textbf{CQ and non-novice programmers}}
CQ measures the difficulty of writing valid programs in  
programming language $\pl$ without knowing anything other than $\pl$'s syntax.
As analyzed in the previous section, the CQ of a language $\pl$ is affected by 
the semantic complexity and flexibility of $\pl$'s features, and their prevalence
 in the program space. Put differently, CQ is influenced by how hard it is for a randomly 
generated program to use a given feature correctly. We argue that there is a connection  
between CQ and the semantic complexity experienced by non-novice programmers.
}

A programmer's task is to write programs that compile. For the programmer, this 
implies producing programs that are both syntactically and semantically valid.    
While syntax is mostly a barrier for novices, we argue that CQ measures a part 
of the semantic complexity experienced by non-novice programmers. Intuitively, 
if frequently-used features, such as variable declarations, are likely to compile when 
generated randomly, then it also takes low effort for the programmer to get them 
right. The programmer can then concentrate on other things instead. Conversely, 
if frequently-used features are unlikely to compile at random, this 
indicates higher effort for the programmer. CQ can help language designers evaluate 
a feature's effects before its release.

{\parindent0pt
\paragraph{\textbf{Long-run adoption of programming languages}} 
Firstly, it is striking that the four most popular languages of the
TIOBE index are all placed in the top half of the CQ ranking. This could  
suggest that languages with high CQ are more likely to become and remain  
popular. Secondly, newer languages have lower CQs than older languages. 
This could be explained by the recent trend towards feature-rich and type-safe 
programming languages, increasing language complexity but decreasing CQ. 
Particularly, Rust with a CQ of almost zero, is known for having a steep learning 
curve with a significant proportion of learners abandoning it because of its
difficulty~\cite{rust-lang-survey}.
}

{\parindent0pt
\paragraph{\textbf{Implications on compiler testing}} 
Another striking finding is C's much higher CQ compared to all other languages. We argue that this could partially explain the  
success of testing campaigns for C compilers ~\cite{yang-etal-pldi2011,le-afshari-pldi2014}.   
Since C compilers have higher CQ \ie, are more permissive, they are also more accessible 
for testers. Conversely, compilers of low CQ languages such as Haskell and Rust are more difficult to test.   
This is especially plausible since many testing approaches for compilers 
implement the language grammars programmatically. As CQ can be a proxy for the 
effectiveness of fuzzers for a given
language, language designers can use \toolname{} to optimize  
for high CQ.}

{\parindent0pt
\paragraph{\textbf{Limitations}} 
Our tool \toolname{} enables a large-scale study of the CQs of 12 popular compiled programming languages.     
However, we acknowledge that our work also comes with limitations. CQ is defined 
using the proportion of compiling programs to the total number of syntactically 
valid programs. One limitation of our work is that we measured CQ    
only on programs with a single entry point and forbade use of the language's standard library. 
However, this ignores other important features contributing to a language's
complexity such as includes in C/C++ and traits in Rust. 
Moreover, CQ does not consider 
the complexity of the language's runtime semantics.}

\section{Related Work}
\label{sec:related-work}
We survey three strands of related work on (1) automated program generation, (2) 
property-based testing, and (3) empirical studies on programming language learning.      

{\parindent0pt 
\paragraph{\textbf{Automated program generation}}
The earliest work laying the ground for automated program generation is Lisp's 
meta-programming, manipulating source code as a data structure by John McCarthy. 
Approaches in program synthesis are also related to our approach because these also 
use program enumeration, \eg, FlashFill~\cite{gulwani-popl2011}.
However, their goals are different. Synthesis aims to find a program satisfying a 
specification. Our approach generates programs to compute CQs.
Genetic programming \cite{koza1992genetic} 
is another approach for program generation from progressively unfit-to-fit programs 
using selection, crossover, replication, and mutation strategies on the program's 
ASTs. Another closely related line of research is \citeauthor{grygiel-lescanne-2013}~[\citeyear{grygiel-lescanne-2013}]'s 
work on counting terms of the lambda calculus, combinatorial properties, and 
asymptotic distribution on random generation. They observe that lambda calculus's 
CQ converges to zero.  Similarly, in our work, all languages show this behaviour. 
The only exception is C for which even many sizeable programs compile.     
}

{\parindent0pt 
\paragraph{\textbf{Property-based testers and grammar-based fuzzers}}
Our work is based on the exhaustive enumerator FEAT~\cite{duregard-etal-2012}, 
which belongs to the family of property-based testers such as 
LeanCheck by Rudy Matela, SmallCheck~\cite{colin-etal-haskell2008}, 
and QuickCheck~\cite{claessen-hughes-icfp00}. Different from property-based 
testers whose test drivers are usually manually written, our approach    
generates test drivers from a context-free grammar automatically. Grammar-based 
black-box fuzzers~\cite{burkhardt-1967,hanford1970,yang-etal-pldi2011} are all 
also technically related to our approach. However, our work and fuzzers differ   
in their objective. The objective of a fuzzer is to find bugs, while 
we aim to measure the compilation quotient.}

{\parindent0pt
\paragraph{\textbf{Empirical studies on programming languages}}
Our study is related to studies on programming languages. \citeauthor{ray-etal-fse2014}~[\citeyear{ray-etal-fse2014}] 
study how programming languages affect software quality from large datasets based 
on GitHub users while other work studies key factors in adopting programming languages 
 with user surveys \cite{meyerovich-etal-oopsla2013}. More loosely related is work 
on how people learn computer programming~\cite{guo2013online} and studies that compare 
the cognitive load of Python programming against programming in a visual programming 
language such as Algot \cite{thorgeirsson-etal-sigcse2024}.
}

\section{Conclusion and Future Work}
\label{sec:conclusion}

We introduced the \emph{compilation quotient} (CQ), a metric for the compilation 
hardness of compiled programming languages. The key idea is measuring the CQs 
of programming languages by sampling programs from context-free grammars 
using a bucket-based sampling algorithm. With our framework \toolname{}, 
we computed the CQs of twelve popular compiled programming languages. 
Our findings show high variation: the language with the 
highest CQ is C  with a CQ of almost 50 leading by a large margin over all other languages.
Strikingly, Rust's CQ is almost zero. 
For all languages, the number of observed compiling programs drops to zero within the 
measured size range, but C is an exception, showing a high number of compiling
programs even at large sizes. Long programs are dominated by either declarations 
or expressions -- CQ is then significantly 
affected by how much the language restricts these constructs. 
CQ is related to extrinsic properties of programming languages -- the most popular languages 
have a high CQ (\eg, C++, Java, C\#), while newer languages have lower CQs (\eg Rust).
We believe that CQ is a useful metric for the comparison and analysis of programming 
languages, providing valuable information to designers of future languages.
By providing a framework for more objective discussions of language complexity, 
CQ can help programmers to make more informed decisions.

We outline three avenues of future work. First, we aim to extend CQ 
to interpreted languages. Unlike in compiled languages where ahead-of-time 
compilers reject invalid programs with errors, interpreters' errors  
could be either caused by invalidity (\eg type errors, incorrect declarations, \etc) or 
runtime issues (\eg division by zero, null reference, \etc). Hence, a key challenge in extending the CQ 
to interpreted languages is to categorize interpreter errors. Another direction 
of future work is to generate more complex programs \eg, with more than a 
single function. Supporting the sampling of complex programs requires 
that the sampled program declare identifiers before their first use. Otherwise, 
the chances of generating valid programs are very small.  Finally, it might be worthwhile to investigate 
how difficult it is to fix errors in different programming languages. One possible 
way to do this is to present each rejected program to a large language model, ask the model to fix 
the error, and then check whether the fixed program compiles.

\balance
\bibliographystyle{ACM-Reference-Format}

\end{document}